\documentclass[aip,
               jcp,
               preprint,
               amsmath,
               superscriptaddress,
               floatfix,
               nobibnotes,
               ]{revtex4-2}

\usepackage{amssymb}
\usepackage{amsmath}
\usepackage{mathtools}
\usepackage{latexsym}
\usepackage{amsfonts}
\usepackage{bm}
\usepackage[dvips]{graphicx}
\usepackage{color}
\usepackage[utf8]{inputenc}
\usepackage[flushleft]{caption}
\usepackage{subfigure}
\usepackage{rotating}
\usepackage{txfonts}
\usepackage{booktabs}
\usepackage[unicode]{hyperref}
\usepackage{algorithm}
\usepackage{algorithmicx}
\usepackage{algpseudocode}
\DeclareMathAlphabet{\mathpzc}{OT1}{pzc}{m}{it}

\def\figfoot{OH + HO$_2$, MRCI}

\newcommand{\figcaption}[2]{
    \noindent {\bf Figure \ref{#1}:} #2
    \vspace{1cm}
}

\begin{document}


\title{%
A Multi-Electronic-State Model to Interpret the Apparent Anomalous
Arrhenius Curve of OH + HO$_2$ $\to$ O$_2$ + H$_2$O}%

\author{Xingyu Zhang}
  \affiliation{Department of Chemistry, 
               Northwestern Polytechnical University,
               West Youyi Road 127, 710072 Xi'an,
               China}

\author{Jinke Yu}
  \affiliation{Department of Chemistry, 
               Northwestern Polytechnical University,
               West Youyi Road 127, 710072 Xi'an,
               China}
               
\author{Qingyong Meng}
 \email{qingyong.meng@nwpu.edu.cn}
  \affiliation{Department of Chemistry, 
               Northwestern Polytechnical University,
               West Youyi Road 127, 710072 Xi'an,
               China}

\date{\today}


\begin{abstract}

{\bf Abstract}:
A comprehensive multi-electronic-state model for OH + HO$_2$
$\to$ O$_2$ + H$_2$O has been developed through extensive multi-reference
configuration interaction (MRCI) calculations, aiming to elucidate two
key experimental observations: (1) an unusually deep and narrow ``well''
in the Arrhenius curve near $1100\:\mathrm{K}$ and (2) a slightly
negative temperature dependence in the range of $200\:\mathrm{K}\sim
500\:\mathrm{K}$. Moreover, the present model can serve as the
basis for constructing multi-state Hamiltonian in multi-dimensional
quantum dynamics calculations. The present model incorporates eight
state-to-state processes involving OH ($\tilde{X}\;{}^2\Pi$) + HO$_2$
($\tilde{X}\;{}^2A''/\tilde{A}\;{}^2A'$), where three of four processes
associated with HO$_2$ ($\tilde{X}\;{}^2A''$) are exothermic, while
those associated with HO$_2$ ($\tilde{A}\;{}^2A'$) have three endothermic
channels with the smallest barrier of 0.107 eV. At temperatures below
$500\:\mathrm{K}$, the processes of HO$_2$ ($\tilde{A}\;{}^2A'$) remain
inaccessible, and the dominance of exothermic pathways results in a
temperature-independent rate constant. To enable $\mathrm{HO}_2$
excitation at temperatures above $900\:\mathrm{K}$, a black-body
radiation model that facilitates endothermic processes is introduced
and reproduces the reduction factor of $0.3711$ at $1100\:\mathrm{K}$.
This value falls within the experimentally observed range of $0.30\sim0.76$.
Furthermore, a recombination process involving HO$_2$ ($\tilde{A}\;{}^2A'$)
is proposed to provide additional reduction of the rate constant. In
conjunction with our previous works ({\it J. Chem. Phys.} {\bf 152}
(2020), 134309 and {\it J. Chem. Theory Comput.} {\bf 20} (2024), 597),
predicted values of the rate constants are well agree with experiments.
At elevated temperatures exceeding $1242\:\mathrm{K}$ (namely $>0.107\:\mathrm{eV}$),
the overall rate constant becomes temperature-dependent due to the
activation of endothermic processes, leading to a distinct well in
the Arrhenius curve between $900\:\mathrm{K}$ and $1242\:\mathrm{K}$.
This is in agreement with the experimental range of
$900\:\mathrm{K}\sim1250\:\mathrm{K}$. 
{\it Excited State}; {\it Arrhenius Curves}
\end{abstract}

\maketitle
%

\section{Introduction\label{sec:intro}}

Kinetics of the OH + HO$_2$ system, known as HO$_x$, plays an important
role in combustion chemistry
\cite{gar00,kon15:3755,li21:105165,zha21:5799,kli22:111975,li22:112093,fer22:619,bei23:112498}.
Over the past decades, extensive observations
\cite{dra87:471,key88:1193,atk89:881,bau05:757,jpl11,kap02:4392,sri06:6602,hon10:5520,zha13:7381,bur13:1540,hon13:565,bur13:547,pal18:4478,zha18:8152,liu19:12667,son20:134309}
for the OH + HO$_2$ $\to$ O$_2$ + H$_2$O reaction have found a deep and
narrow ``well'' in the temperature-dependent Arrhenius curve near
$1100\:\mathrm{K}$ and a slightly negative temperature-dependence at
the range of $200\:\mathrm{K}\sim500\:\mathrm{K}$, as illustrated by
Figure \ref{fig:expt-arr-plot}. According to International Union of
Pure and Applied Chemistry (IUPAC) \cite{atk89:881,bau05:757} and Jet
Propulsion Laboratory (JPL) \cite{jpl11}, together with the results of
Keyser \cite{key88:1193}, disregarding the very low values, reported
rate constants at low temperature range from $250\:\mathrm{K}$ to
$500\:\mathrm{K}$ fall into two groups. One set of lower values at
$\sim7\times10^{-11}\:\mathrm{cm}^3\cdot\mathrm{molecule}^{-1}\cdot\mathrm{s}^{-1}$
was obtained at low pressures. Another set of higher values at
$\sim1\times10^{-10}\:\mathrm{cm}^3\cdot\mathrm{molecule}^{-1}\cdot\mathrm{s}^{-1}$
was obtained at pressures close to one bar. Dransfeld {\it et al.}
\cite{dra87:471} suggested the rate constant at room temperature is
pressure dependent. Later, Keyser \cite{key88:1193} suggested that
the discrepancy between these two sets of values was due to secondary
processes of small amounts of H and O atoms in the low-pressure case,
rather than pressure dependence. Moreover, negative temperature-dependence
(the black solid line in Figure \ref{fig:expt-arr-plot}) of
$4.8\times10^{-11}\cdot\exp(250/T)$ found by Keyser \cite{key88:1193}
was recommended by both IUPAC \cite{atk89:881,bau05:757} and JPL
\cite{jpl11}. The only other determination of the temperature dependence
(the yellow solid line in Figure \ref{fig:expt-arr-plot}) supports such
small negative dependence even though the absolute values appear to be
systematically low. However, as shown by Figure \ref{fig:expt-arr-plot},
extrapolation of such temperature dependence clearly does not reproduce
the available data at higher temperature above $500\:\mathrm{K}$. A
shock tube study over the temperature range from $1118\:\mathrm{K}$ to
$1566\:\mathrm{K}$ (circle symbols $\circ$ in Figure \ref{fig:expt-arr-plot})
suggested that the rate constant initially declines in value as the
temperature increases reaching a minimum at $\sim1250\:\mathrm{K}$
and rises again at temperatures beyond. This behavior was suggestive
of formation and decomposition of an intermediate complex, with the
implication that the rate constant is possibly pressure dependence.
The preferred values at high temperatures was suggeted \cite{atk89:881,bau05:757}
to be $8.3\times10^{-11}\:\mathrm{cm}^3\cdot\mathrm{molecule}^{-1}\cdot\mathrm{s}^{-1}$,
$1.0\times10^{-10}\:\mathrm{cm}^3\cdot\mathrm{molecule}^{-1}\cdot\mathrm{s}^{-1}$,
and $3.3\times10^{-11}\:\mathrm{cm}^3\cdot\mathrm{molecule}^{-1}\cdot\mathrm{s}^{-1}$.
In 2013, Hong {\it et al.} \cite{hon13:565} reported a set of rather
low values (diamond symbols $\Diamond$ in Figure \ref{fig:expt-arr-plot})
by shock tube in conjunction with multi-species laser techniques. The
above experimental observations are collected in the Supporting
Information. Unless otherwise specified, in this work, values of the
rate constants will be all given in $\mathrm{cm}^3\cdot\mathrm{molecule}^{-1}\cdot\mathrm{s}^{-1}$.
In order to explain such apparent anomalous Arrhenius curve, a
multi-electronic-state model is proposed in this work by
determining state-to-state processes through extensive multi-reference
configuration interaction (MRCI) calculations. By the present model,
an interpretation is given to understand the above previously observed
experiments, together with multi-scale simulations reported by Burke
and co-workers \cite{bur13:547}.

Over the past years, to interpret the above characteristic, various
approaches were employed to build the potential energy surface (PES)
based on which extensive dynamics simulations were launched leading
to fruitful understandings on the OH + HO$_2$ $\to$ O$_2$ + H$_2$O
reaction. With the aid of the coupled-cluster singles, doubles, and
perturbative triples (CCSD(T)) energy calculations, the permutation
invariant polynomial (PIP) neutral-network (NN) \cite{liu19:12667}
and Gaussian process regression (GPR) \cite{son20:134309} methods were
used to construct singlet-state PES. In 2019, on the basis of their
PIP-NN PES, Li and co-workers \cite{liu19:12667} performed extensive
quasi-classical trajectory (QCT) calculations to compute the rate
constants. In 2020, on the GPR PES \cite{son20:134309} ring-polymer
molecular dynamics (RPMD) calculations \cite{son20:134309} predicted
rate constants closed to the previous ones \cite{liu19:12667}. To
interpret temperature-dependence of the rate constant, a simple
kinetics model \cite{men18:8320,son20:134309} was proposed. Later, on
the basis of the PIP-NN PES as well as the RPMD method, Guo and co-workers
\cite{liu20:3331} further compute the rate constants that are close to
the previous RPMD results based on the GPR PES \cite{son20:134309}. In
2024, a Jacobi-like coordinates set \cite{son24:597} was employed to
derive the kinetic energy operator (KEO) by polyspherical approach
\cite{gat09:1}; meanwhile, a two-state nonadiabatic PES \cite{son24:597}
was constructed through the training database computed at the MRCI level.
Having obtained such two-state nonadiabatic Hamiltonian operator, multi-layer
vertion \cite{wan03:1289,man08:164116,ven11:044135,wan15:7951} of multi-configurational
time-dependent Hartree (MCTDH), called ML-MCTDH, was employed to propagate
the nuclear wave function and to compute energy-dependent reaction probability.
However, a clear explanation for the unusual temperature-dependence is still
missed by the above studies.

The rest of this paper is organized as follows. In Section \ref{sec:theore-work},
we will describe the present calculation details. Section \ref{sec:results}
presents model by the MRCI results and gives discussions and perspectives.
Finally, Section \ref{sec:con} concludes with a summary.

\section{Theoretical Framework\label{sec:theore-work}}

\begin{figure}
  \centering
   \includegraphics[width=18cm]{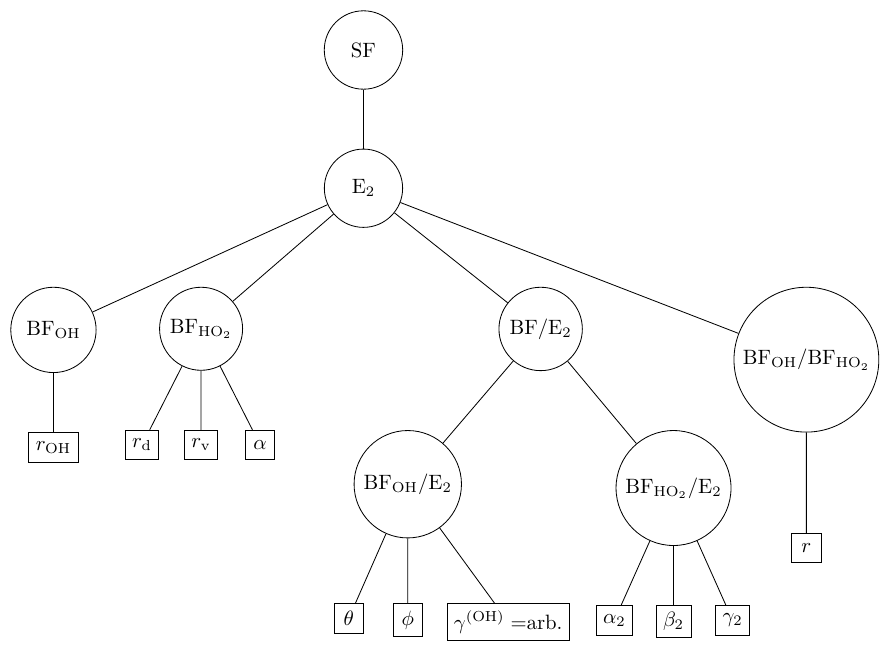}
    \caption{\figfoot}
     \label{fig:MLtree}
      \end{figure}

The OH + HO$_2$ system is a three-molecular-channels system composed
by two primitive reactions
\begin{align}
\mathrm{OH}+\mathrm{HO}_2&\to\mathrm{O}_2+\mathrm{H}_2\mathrm{O},\quad\Delta E=-3.035\;\mathrm{eV},
\label{eq:o2-wat} \allowdisplaybreaks[4]  \\
\mathrm{OH}+\mathrm{HO}_2&\to\mathrm{H}_2+\mathrm{O}_3,\quad\Delta E=1.623\;\mathrm{eV},
\label{eq:h2-o3}
\end{align}
together with other secondary and minor reactions, say
\begin{equation}
\mathrm{HO}_2\to\mathrm{H}+\mathrm{O}_2,\;
\mathrm{O}_3\to\mathrm{O}+\mathrm{O}_2,\;
\mathrm{H}_2\mathrm{O}\to\mathrm{O}+\mathrm{OH},\;
\mathrm{OH}\to\mathrm{H}+\mathrm{O},\;
\mathrm{O}_2\to\mathrm{O}+\mathrm{O}.
\label{eq:secondary-reacx} 
\end{equation}
The $\Delta E$ values are energy differences which indicate that 
Reaction \eqref{eq:o2-wat} is most likely to occur. In addition,
there exists another reaction involved oxygen atom,
\begin{equation}
\mathrm{OH}+\mathrm{HO}_2\to\mathrm{O}+\mathrm{H}_2\mathrm{O}_2,
\quad\Delta E=0.811\;\mathrm{eV},
\label{eq:o-h2o2}
\end{equation}
which is also unlikely occurs due to $\Delta E>0$. Therefore, main
purpose of the present work concerns on the OH + HO$_2$ $\to$ O$_2$
+ H$_2$O reaction. As illustrated by Figure \ref{fig:MLtree}, in the
present calculations the HO$_x$ system is described by Jacobi-like
coordinates which were proposed previously \cite{son24:597}. The OH
and HO$_2$ fragments are represented by $\mathrm{BF}_{\mathrm{OH}}$
and $\mathrm{BF}_{\mathrm{HO}_2}$, respectively, while the E$_2$ system
is used to represent the whole system. The relative vector from the
center-of-mass (COM) of HO$_2$ to that of OH is $\vec{r}$, while
$\vec{r}_{\mathrm{OH}}$ and $\vec{r}_{\mathrm{v}}$ are vectors along
the O-H and O-O bonds, respectively. The vector from the COM of O$_2$
to the H atom is $\vec{r}_{\mathrm{d}}$. As discussed previously, this
representation is suitable for subsequent quantum dynamics calculations.
Therefore, the present MRCI results can be directly used in building
the potential energy surface (PES) which is required for dynamics
calculations.

Given in Table \ref{tab:mrci-details} are numerical details of the
present MRCI calculations. All of the present multi-reference (MR)
{\it ab initio} calculations with full active space and single-reference
(SR) {\it ab initio} are carried out by the MOLPRO package \cite{wer12:242,molpro}.
With a quadratically convergent multi-configurational self-consistent
field (MCSCF) \cite{wer85:5053,kno85:259} wave function constituting
the reference function, the internally contracted single and double
excitation MRCI, called ic-MRCISD \cite{kno84:315,wer85:5053}, is
performed to approximately compute the MRCI wave function and energy.
The SR {\it ab initio} calculations are performed to optimize the
fragments geometries and reproduce the ground-state electronic wave
function to determine the active space. Moreover,
the augmented Dunning's correlation consistent polarized valence triple
zeta, {\it i.e.} aug-cc-pVTZ, \cite{dun89:1007,ken92:6796} basis set is
used. Since MCSCF and MRCI are implemented with Abelian groups, calculations
for high-symmetry fragments must be performed in subgroup. For O$_2$
and H$_2$, the MR calculations are performed in the $D_{2\mathrm{h}}$
subgroup of $D_{\infty\mathrm{h}}$, where $\Sigma_g^+$ corresponds to
the $A_g$ irreducible representation, $\Sigma_u^+$ to $B_{1u}$,
$\Sigma_g^-$ to $B_{1g}$, $\Pi_g$ to $B_{2g}\oplus B_{3g}$, $\Pi_u$ to
$B_{2u}\oplus B_{3u}$, $\Delta_g$ to $A_g\oplus B_{1g}$, and $\Delta_u$
to $A_u\oplus B_{1u}$. For OH, the $C_{\infty\mathrm{v}}$ group is
replaced by the $C_{2\mathrm{v}}$ subgroup, where $\Sigma^+$ corresponds
to $A_1$, $\Sigma^-$ to $A_2$, and $\Pi$ to $B_1\oplus B_2$. Calculations
for HO$_2$ are performed in $C_{\mathrm{s}}$, while H$_2$O and O$_3$
in $C_{2\mathrm{v}}$.

At the CCSD(T)/aug-cc-pVTZ level, geometries of the fragments at the
ground states are firstly optimized to predict Reactions \eqref{eq:o2-wat},
\eqref{eq:h2-o3}, and \eqref{eq:o-h2o2}. Analysis of the CCSD(T) wave
function predicts informations to launch the MRCI calculations for each
fragment. Table \ref{tab:mrci-details} gives configuration of the
ground state and its reference coefficient, together with the active
space. As expected, all of fragments at the ground state are primary
where the valence orbitals are occupied without ``hole'' in the inner
orbitals. As given in Table \ref{tab:mrci-details}, the active space
with all of valence orbitals and valence electrons is used for each
fragment in this work. Next, the vertical excitation energies for
each fragment are computed at the MRCI/aug-cc-pVTZ level to determine
electronic states. Geometries and total energies of these states are
then optimized at the same level leading to the adiabatic excitation
energies. By adiabatic energies, one can obtain lower-lying electronic
states of each channel and, unless otherwise specified in this work,
set energy of the lowest-lying fragment O$_2$ ($\tilde{X}\;{}^3\Sigma_g^-$) + H$_2$O
($\tilde{X}\;{}^1A_1$) to be zero in this work.

With adiabatic excitation energies of all three channels, one can obtain
energy correlation diagram allowing us understanding the state-state
potential energy curves (PECs) along reaction coordinates and interpreting
static mechanisms. Letting $\Gamma_{\mathrm{A}}$ and $s_{\mathrm{A}}$
be state symmetry ({\it i.e.} irreducible representation of the state)
and spin angular-momentum of the A fragment, respectively, the A + B
channel belongs to the ${}^{2s+1}\Gamma={}^{2s+1}(\Gamma_\mathrm{A}\otimes\Gamma_{\mathrm{B}})$
states, where $s=\vert s_{\mathrm{A}}-s_{\mathrm{B}}\vert,\vert
s_{\mathrm{A}}-s_{\mathrm{B}}\vert+1,\cdots,\vert s_{\mathrm{A}}+
s_{\mathrm{B}}\vert$. Due to the non-crossing rule of the PECs at the
same symmetry and the conservation relations of total spin, one can
connect two channels at appropriate states leading to correlation
diagrams for Reactions \eqref{eq:o2-wat} and \eqref{eq:h2-o3}. Correlation
diagrams also illustrate crossing points of two PECs which implies
minimum energy crossing points (MECPs). Due to the coupling terms of
the potential matrix, crossing points play an important role in
nonadiabatic mechanism. Due to the unavailability of analytical MRCI
energy gradients, geometries of the present MECPs are optimized at the
MCSCF/aug-cc-pVTZ level, while their energies are computed at the
MRCI//MCSCF level.

\section{Results\label{sec:results}}

\subsection{State-to-State Scattering Processes\label{sec:dynamics}}

\begin{figure}
  \centering
   \includegraphics[width=18cm]{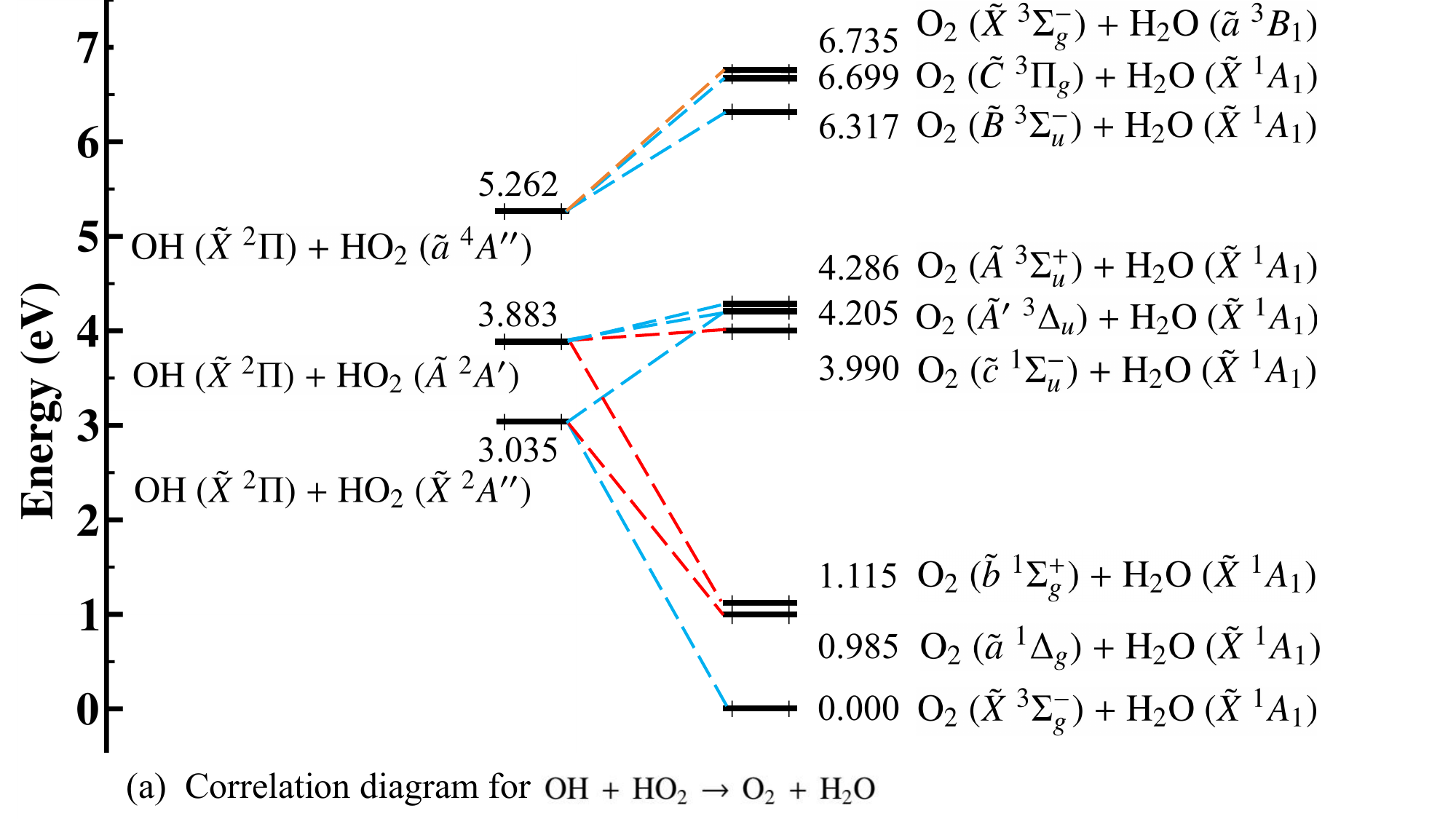}
    \includegraphics[width=18cm]{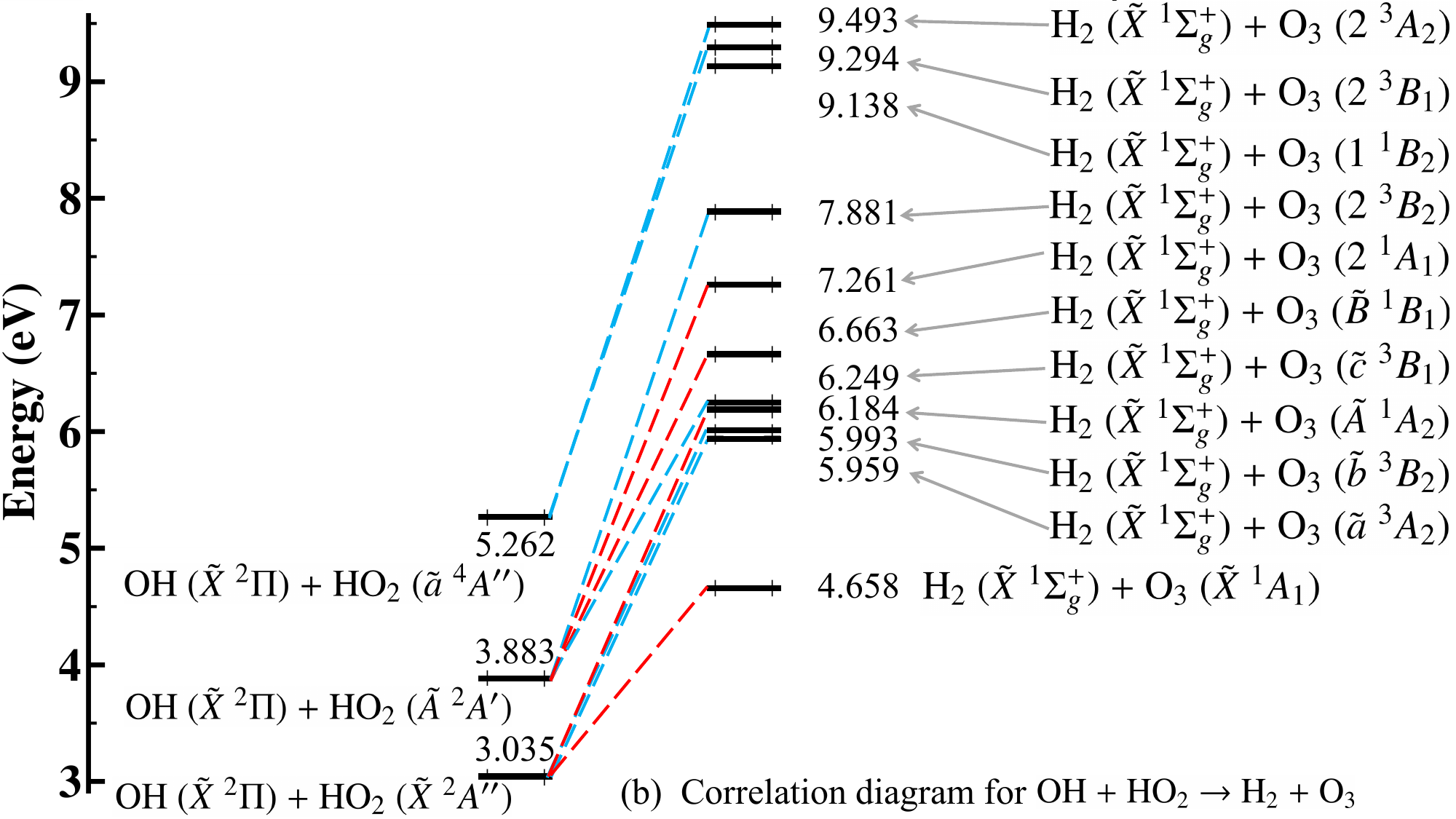}
    \caption{\figfoot}
     \label{fig:corr-diag}
      \end{figure}

By the present MRCI geometry optimizations for fragments, we collect
all probable electronic states of reactant channel, {\it i.e.} OH +
HO$_2$, and two product channels, {\it i.e.} O$_2$ + H$_2$O and H$_2$
+ O$_3$, in the Supporting Information. Optimized geometry parameters,
relative energy values, and irreducible representations of electronic
states are also given there. With these state-resolved relative energies
and symmetries, one can obtain correlation diagrams of Reactions
\eqref{eq:o2-wat} and \eqref{eq:h2-o3} as shown in Figure \ref{fig:corr-diag},
where relative energy values are also given. For the OH + HO$_2$ and
O$_2$ + H$_2$O channels, the lower-lying states below $6.800$ eV are
given, while the lower-lying states below $9.500$ eV are given for the
H$_2$ + O$_3$ channel. At the MRCI level, all of state-to-state processes
are found to be barrierless by inspecting their PECs. Therefore, the
correlation diagrams shown in Figure \ref{fig:corr-diag} can illustrate
energy profiles of these processes. By these MRCI calculations, the
following three points can be found.

First, the correlation diagram given in Figure \ref{fig:corr-diag}(a)
indicates that Reaction \eqref{eq:o2-wat} at the ground state is
dramatically exothermic with energy difference of $-3.035$ eV,
\begin{equation}
\mathrm{OH}\;(\tilde{X}\;{}^2\Pi)+\mathrm{HO}_2\;(\tilde{X}\;{}^2A'')\to
\mathrm{O}_2\;(\tilde{X}\;{}^3\Sigma_g^-)+\mathrm{H}_2\mathrm{O}\;
(\tilde{X}\;{}^1A_1),\quad\Delta E=-3.035\;\mathrm{eV}.
\label{eq:o2-wat-XXXX}
\end{equation}
As shown by Figure \ref{fig:corr-diag}(b), Reaction \eqref{eq:h2-o3}
at the ground state is endothermic,
\begin{equation}
\mathrm{OH}\;(\tilde{X}\;{}^2\Pi)+\mathrm{HO}_2\;(\tilde{X}\;{}^2A'')\to
\mathrm{H}_2\;(\tilde{X}\;{}^1\Sigma_g^+)+\mathrm{O}_3\;(\tilde{X}\;{}^1A_1),
\quad\Delta E=1.623\;\mathrm{eV}.
\label{eq:o3-hyd-XXXX}
\end{equation}
In this context, Reaction \eqref{eq:o2-wat} is much more likely to occur
than Reaction \eqref{eq:h2-o3}. Moreover, the correlation diagram of
Reaction \eqref{eq:o2-wat} in Figure \ref{fig:corr-diag}(a) indicates
that scattering processes at lower-lying excited states below $3.900$
eV are exothermic with $\Delta E$ of $-2.050$ and $-2.768$ eV, while
those at higher-lying excited states above $3.800$ eV are endothermic
with $\Delta E$ in the range from $0.107$ eV to $1.437$ eV. For Reaction
\eqref{eq:h2-o3}, scattering processes at excited states below $9.500$
eV are all extremely endothermic with $\Delta E\geq2.780$ eV, as shown
in Figure \ref{fig:corr-diag}(b). Therefore, at lower-lying energy region,
we focus on Reaction \eqref{eq:o2-wat} rather than Reaction \eqref{eq:h2-o3}
to interpret the experiments on the HO$_x$ system.

Second, we focus on the correlation diagram of Reaction \eqref{eq:o2-wat}
in Figure \ref{fig:corr-diag}(a) and refer the reader to the Supporting
Information for details of the electronic states of each fragments.
Since OH ($\tilde{X}\;{}^2\Pi$) and HO$_2$ ($\tilde{X}\;{}^2A''$) are
doubly and singly degenerated, respectively, the reactant channel at
the ground state must be doubly degenerated corresponding to the first
two ${}^{1,3}A$ states of the HO$_x$ system. Due to the non-crossing
rule of two adiabatic PESs in identical symmetry and conservation of
spin, the OH ($\tilde{X}\;{}^2\Pi$) + HO$_2$ ($\tilde{X}\;{}^2A''$)
channel at singlet state connects the product channel O$_2$
($\tilde{a}\;{}^1\Delta_g$) + H$_2$O ($\tilde{X}\;{}^1A_1$) which is
also doubly degenerated,
\begin{equation}
\mathrm{OH}\;(\tilde{X}\;{}^2\Pi)+\mathrm{HO}_2\;(\tilde{X}\;{}^2A'')
\to\mathrm{O}_2\;(\tilde{a}\;{}^1\Delta_g)+\mathrm{H}_2\mathrm{O}\;
(\tilde{X}\;{}^1A_1),\quad\Delta E=-2.050\;\mathrm{eV}.
\label{eq:o2-wat-XXaX}
\end{equation}
Thus, the lowest-lying two singlet states of the HO$_x$ system are
degenerated, or at least approximately so, at the region near the
reaction coordinate. At the first glance, this should be an example of
accidental degeneracy because the HO$_x$ system belongs to $C_1$. However,
the double degeneracy of Reaction \eqref{eq:o2-wat-XXaX} is comprehensible
because the 13th and 14th molecular orbitals (MOs) of HO$_x$ are nearly
degenerated. It is these two MOs that constitute the $1\pi$ orbital of
OH and the $1\pi_g$ orbital of O$_2$ which are both doubly degenerated.
However, the situation changes for the lowest-lying two triplet states
where the OH ($\tilde{X}\;{}^2\Pi$) + HO$_2$ ($\tilde{X}\;{}^2A''$)
channel is divided into two branches during reaction. The first one is
the ground-state product channel, O$_2$ ($\tilde{X}\;{}^3\Sigma_g^-$) +
H$_2$O ($\tilde{X}\;{}^1A_1$), which is singly degenerated. The other
branch is O$_2$ ($\tilde{A}'\;{}^3\Delta_u$) + H$_2$O ($\tilde{X}\;{}^1A_1$).
Thus, the triplet states of OH ($\tilde{X}\;{}^2\Pi$) + HO$_2$
($\tilde{X}\;{}^2A''$) lead to exothermic Reaction \eqref{eq:o2-wat-XXXX}
and endothermic process,
\begin{equation}
\mathrm{OH}\;(\tilde{X}\;{}^2\Pi)+\mathrm{HO}_2\;(\tilde{X}\;{}^2A'')
\to\mathrm{O}_2\;(\tilde{A}'\;{}^3\Delta_u)+\mathrm{H}_2\mathrm{O}\;
(\tilde{X}\;{}^1A_1),\quad\Delta E=1.170\;\mathrm{eV}.
\label{eq:o2-wat-XXApX}
\end{equation}
Similarly, one can analyze processes from OH ($\tilde{X}\;{}^2\Pi$) +
HO$_2$ ($\tilde{A}\;{}^2A'$), which is related with four states, leading
to
\begin{align}
\mathrm{OH}\;(\tilde{X}\;{}^2\Pi)+\mathrm{HO}_2\;(\tilde{A}\;{}^2A')&\to
\mathrm{O}_2\;(\tilde{b}\;{}^1\Sigma_g^+)+\mathrm{H}_2\mathrm{O}\;
(\tilde{X}\;{}^1A_1),\quad\Delta E=-2.768\;\mathrm{eV},
\label{eq:o2-wat-XAbX} \allowdisplaybreaks[4] \\
\mathrm{OH}\;(\tilde{X}\;{}^2\Pi)+\mathrm{HO}_2\;(\tilde{A}\;{}^2A')&\to
\mathrm{O}_2\;(\tilde{c}\;{}^1\Sigma_u^-)+\mathrm{H}_2\mathrm{O}\;
(\tilde{X}\;{}^1A_1),\quad\Delta E=0.107\;\mathrm{eV},
\label{eq:o2-wat-XAcX} \allowdisplaybreaks[4] \\
\mathrm{OH}\;(\tilde{X}\;{}^2\Pi)+\mathrm{HO}_2\;(\tilde{A}\;{}^2A')&\to
\mathrm{O}_2\;(\tilde{A}'\;{}^3\Delta_u)+\mathrm{H}_2\mathrm{O}\;
(\tilde{X}\;{}^1A_1),\quad\Delta E=0.322\;\mathrm{eV},
\label{eq:o2-wat-XAApX} \allowdisplaybreaks[4] \\
\mathrm{OH}\;(\tilde{X}\;{}^2\Pi)+\mathrm{HO}_2\;(\tilde{A}\;{}^2A')&\to
\mathrm{O}_2\;(\tilde{A}\;{}^3\Sigma_u^+)+\mathrm{H}_2\mathrm{O}\;
(\tilde{X}\;{}^1A_1),\quad\Delta E=0.403\;\mathrm{eV}.
\label{eq:o2-wat-XAAX}
\end{align}
These processes are all singly degenerated, while Reactions \eqref{eq:o2-wat-XXApX}
and \eqref{eq:o2-wat-XAApX} share the same product channel due to doubly
degenerated O$_2$ ($\tilde{A}'\;{}^3\Delta_u$) fragment.

Finally, it is necessary to consider processes with the third excited
state of reactant, {\it i.e.} OH ($\tilde{X}\;{}^2\Pi$) + HO$_2$
($\tilde{a}\;{}^4A''$) and to determine a nearly complete Hilbert
subspace for electronic motions. This will be helpful to further
consider non-adiabatic Hamiltonian for quantum dynamics of HO$_x$.
First of all, energy of HO$_2$ ($\tilde{a}\;{}^4A''$) is $2.227$ eV
which is larger than that of HO$_2$ ($\tilde{X}\;{}^2A''$). Such
large value and spin conversion prohibit occurrence of HO$_2$
($\tilde{a}\;{}^4A''$). Moreover, all of processes from OH
($\tilde{X}\;{}^2\Pi$) + HO$_2$ ($\tilde{a}\;{}^4A''$) are dramatically
endothermic with energy differences in the range from $1.055$ eV to
$1.473$ eV (see Figure \ref{fig:corr-diag}(a)). Due to such large values,
these state-to-state processes hardly occur at usual conditions. Second,
geometry optimization for HO$_2$ ($\tilde{a}\;{}^4A''$) predicts rather
large $r_{\mathrm{d}}$ value of $2.958$ {\AA} which implies that HO$_2$
($\tilde{a}\;{}^4A''$) is a weakly bound state and can much more easily
dissociate to H (${}^2S_u$) + O$_2$ ($\tilde{X}\;{}^3\Sigma_g^-$) rather
than react with OH. Thus, processes from HO$_2$ ($\tilde{a}\;{}^4A''$)
are not considered. Third, Figure \ref{fig:corr-diag}(a) explicitly
illustrates that all of lower-lying eight processes are decoupled with
those from HO$_2$ ($\tilde{a}\;{}^4A''$) and thus HO$_2$ ($\tilde{a}\;{}^4A''$)
is unimportant during Reaction \eqref{eq:o2-wat}. In this context, by the
lower-lying eight processes, that form the nearly complete Hilbert subspace
of electronic representation, the multi-state model for Reaction
\eqref{eq:o2-wat} can be proposed to explain temperature-dependence of
the rate constants, as shown by Figure \ref{fig:expt-arr-plot}).

\subsection{Multi-Electronic-State Model\label{sec:kinetics-exp}}

The present multi-state model contains four singlet processes,
including doubly degenerated Reaction \eqref{eq:o2-wat-XXaX} together
with Reactions \eqref{eq:o2-wat-XAbX} and \eqref{eq:o2-wat-XAcX}, and
another four triplet processes, including Reactions \eqref{eq:o2-wat-XXXX},
\eqref{eq:o2-wat-XXApX}, \eqref{eq:o2-wat-XAApX}, and \eqref{eq:o2-wat-XAAX}.
These processes all occur at the energy below $4.3$ eV relative to O$_2$
($\tilde{X}\;{}^3\Sigma_g^-$) + H$_2$O ($\tilde{X}\;{}^1A_1$). According
to Table \ref{tab:mrci-details}, in addition to primary Reaction \eqref{eq:o2-wat},
there exist other secondary processes and those with the least importance.
Moreover, transitions of the reactant fragments should be also introduced.
Since OH always maitains at the gound state and the computed oscillator
strength $f=0.987$ for transition HO$_2$ ($\tilde{A}\;{}^2A'$) $\leftarrow$
HO$_2$ ($\tilde{X}\;{}^2A''$) is not small, transition between the first
two reactant states is allowable. This is not surprising because
$\tilde{A}\;{}^2A'$ and $\tilde{X}\;{}^2A''$ can be coupled with the
aid of a $a''$ vibrational mode which exists for the HO$_2$ fragment.
In addition, Reaction \eqref{eq:o3-hyd-XXXX} and dissociation processes
\begin{align}
\mathrm{HO}_2\;(\tilde{X}\;{}^2A'')&\to\mathrm{H}\;({}^2S_u)+\mathrm{O}_2\;
(\tilde{X}\;{}^3\Sigma_g^-),\quad\Delta E=2.224\;\mathrm{eV}, 
\label{eq:diss-of-ho2-XX}  \allowdisplaybreaks[4] \\
\mathrm{HO}_2\;(\tilde{A}\;{}^2A')&\to\mathrm{H}\;({}^2S_u)+\mathrm{O}_2\;
(\tilde{a}\;{}^1\Delta_g),\quad\Delta E=2.361\;\mathrm{eV}, 
\label{eq:diss-of-ho2-Xa}  \allowdisplaybreaks[4] \\
\mathrm{HO}_2\;(\tilde{a}\;{}^4A'')&\to\mathrm{H}\;({}^2S_u)+\mathrm{O}_2\;
(\tilde{X}\;{}^3\Sigma_g^-),\quad\Delta E=0.179\;\mathrm{eV}, 
\label{eq:diss-of-ho2-aX}
\end{align}
may also be contained in the present model. Their roles will be clear later
in Section \ref{sec:quan-dyn}. With the above considerations, we give the
present multi-state model in Table \ref{tab:net-hox-kin}, including
the above eight state-to-state processes and six secondary processes.

\begin{figure}
  \centering
   \includegraphics[width=18cm]{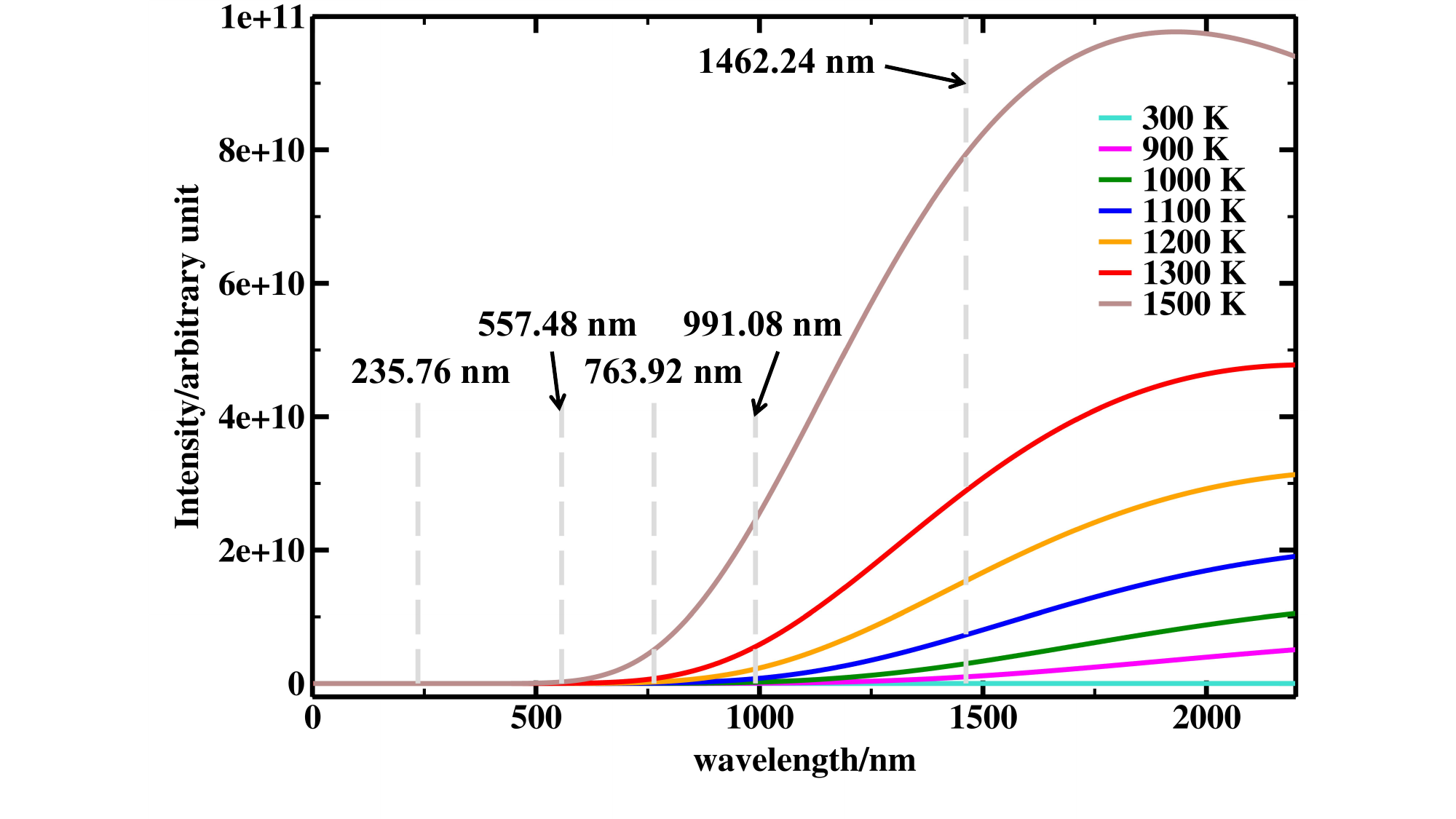}
    \caption{\figfoot}
     \label{fig:blackbody}
      \end{figure}

Before interpreting the Arrhenius curve, we must turn to possibility
of the HO$_2$ ($\tilde{A}\;{}^2A'$) $\leftarrow$ HO$_2$ ($\tilde{X}\;{}^2A''$)
transition as well as energy limitations of the present model. To this
end, the measurement apparatus for kinetics of HO$_x$ is supposed to be
a black body. This assumption is possible because the combustion or
atmosphere system in measurement is a typical thermal-radiation system
and approximately in stable from the energetic viewpoint even though
it is essentially a typical open system. With this assumption, Planck
radiation law predicts the radiation distribution $M$ as a function of
radiation wavelength $\lambda$ and temperature $T$ in the form,
\begin{equation}
M(\lambda,T)=\frac{4\pi^2\hbar c^2}{\lambda^5}\left[
\exp\left(\frac{2\pi\hbar c}{\lambda k_{\mathrm{B}}T}\right)-1\right]^{-1}
=\frac{\gamma}{\lambda^5}\left[\exp\left(\frac{b}{\lambda T}\right)-1\right]^{-1}
\label{eq:radiation-dis-bb}
\end{equation}
Figure \ref{fig:blackbody} illustrates $M(\lambda,T)$ as function of
$\lambda$ at typical temperatures, including $300$ K, $900$ K, $1000$
K, $1100$ K, $1200$ K, $1300$ K, and $1500$ K, together with several
energy (in nm) criteria relative to OH ($\tilde{X}\;{}^2\Pi$) + HO$_2$
($\tilde{X}\;{}^2A''$). The first dissociation energy for H$_2$ + O$_3$
and the largest energy value of the present electronic Hilbert subspace
are related with the wavelength values of $763.92$ nm and $991.08$ nm,
respectively. The former represents energy criterion to open the H$_2$
+ O$_3$ channel, while the latter represents the largest energy value
of the present model. The HO$_2$ excitation energy of transition
$\tilde{A}\;{}^2A'\leftarrow\tilde{X}\;{}^2A''$ is related with the
wavelength of $\lambda_{\mathrm{ex}}=1462.24$ nm which is criterion to
open state-to-state processes of HO$_2$ ($\tilde{A}\;{}^2A'$). Further,
rate constants of endothermic processes depend on temperature by the
Boltzmann factor
\begin{equation}
F_{\mathrm{B}}(T)=\exp\left(-\frac{\Delta E}{k_{\mathrm{B}}T}\right),
\label{eq:tst-ini-fin}
\end{equation}
where $\Delta E$ means energy difference or potential barrier and $k_{\mathrm{B}}$
the Boltzmann constant. By $F_{\mathrm{B}}(T)$, one can estimate difference
of rate constant between low and high temperatures to interpret the
Arrhenius curve. Values of $F_{\mathrm{B}}(T)$ at $1100$ K and $300$ K
are also given in Table \ref{tab:net-hox-kin}.

At low temperatures range from $200\;\mathrm{K}$ to $500\;\mathrm{K}$,
as shown in Figure \ref{fig:blackbody}, distribution of
$\lambda_{\mathrm{ex}}=1462.24$ nm at $300$ K (the light green line)
is nearly equal to zero until temperature approaches to $900$ K (the
purple line) implying that processes with HO$_2$ ($\tilde{A}\;{}^2A'$)
are closed. As given by Table \ref{tab:net-hox-kin}, Reactions \eqref{eq:o2-wat-XXXX}
and \eqref{eq:o2-wat-XXaX} are exothermic without any potential barrier,
while Reaction \eqref{eq:o2-wat-XXApX} is endothermic with the Boltzmann
factor of $\sim10^{-20}$. Thus, contribution of Reaction \eqref{eq:o2-wat-XXApX}
can be ignored which implies that the rate contant of Reaction \eqref{eq:o2-wat}
is temperature-independent. Our previously RPMD calculations \cite{son20:134309}
have predicted rate constants at low temperatures in good agreement
with experimental results. At high tremperature above $900$ K, on the
other hand, the situation changes, where distribution of
$\lambda_{\mathrm{ex}}=1462.24$ nm is much more considerable to open
processes with HO$_2$ ($\tilde{A}\;{}^2A'$) than that at low temperature.
In this context, it is possible that Reaction \eqref{eq:o2-wat} becomes
partially endothermic. According to the mechanism of parallel reactions,
the total rate constant of the overall OH + HO$_2$ $\to$ O$_2$ + H$_2$O
reaction is given by
\begin{equation}
k(T)\sim
k_{\lambda}(T)=\mathcal{Z}(T)k_{\mathrm{exo}}+
\frac{M(\lambda,T)}{M_{\mathrm{tot}}}
\mathcal{Z}(T)F_{\mathrm{B}}(T),
\label{eq:rate-overall}
\end{equation}
where $k_{\mathrm{exo}}$ is total rate constant of exothermic processes,
$M_{\mathrm{tot}}$ normalized constant of the distribution $M(\lambda,T)$,
while $\mathcal{Z}(T)$ is function of partition functions. Considering
weights of endothermic reactions at $1100$ K, the total rate constant
could be reduced by a factor of $0.3711$, that is rate constant at low
temperature should be $\sim2.7$ times larger than that at $1100$ K.
According to Equation \eqref{eq:rate-overall}, this factor is simply
summation of the Boltzmann factors of endothermic processes with
HO$_2$ ($\tilde{A}\;{}^1A'$). This can be understood by (1) noting
that the present model is essentially mechanism of parallel reactions
and (2) assuming that partition functions of the reactant group OH +
HO$_2$ are electronic-state independent. The resulting reduced factor
of $0.3711$ is well agree with experimental results. For example, the
IUPAC \cite{atk89:881,bau05:757} and JPL \cite{jpl11} recommended rate
constant of $\sim1.1\times10^{-10}$ at $300$ K which is
$3.3\sim1.32$ times larger than that of $(3.3\sim8.3)\times10^{-11}$ at
$\sim1100$ K recommended by the same organizes \cite{atk89:881,bau05:757,jpl11}.
One of possible reasons for these uncertainties may be pressure-dependence
of the rate constants, while the other reason might be measurement errors.
Later in Section \ref{sec:quan-dyn}, we will revise the present model
to fix its over-valuation (noting the computed factor of $\sim2.7$ and
the experimental one $3.3\sim1.32$) by discussing on previous experiments
of Keyser \cite{key88:1193}. These experiments suggested that the discrepancy
of rate constants at low temperature is due to secondary processes of small
amounts of H and O atoms in the low-pressure case, rather than pressure
dependence.

Next, the above model must be considered more closely and quantitatively
to interpret temperature range of $900\;\mathrm{K}\sim1250\;\mathrm{K}$
of the well in the Arrhenius curve, called $T_{\mathrm{min}}$. As shown
above, rate constants at the room temperature $T_{\mathrm{room}}$ are
slightly temperature-independent \cite{son20:134309} and $\sim2.7$ times
larger than $k(T=1100\;\mathrm{K})$, while $k(T>900\;\mathrm{K})$ should
increase with increasing temperature. This clearly implies existence of
$T_{\mathrm{min}}$. At the first glance, $T_{\mathrm{min}}$ can be obtained
by calculating the minimum value of $k(T)$. According to Equation \eqref{eq:rate-overall},
change ratio of rate constant relative to that at $T_{\mathrm{room}}=300\;\mathrm{K}$
is given by
\begin{equation}
J_{\lambda}(T)=\frac{\vert k(T)-k(T_{\mathrm{room}})\vert}{k(T_{\mathrm{room}})}
\propto M(\lambda,T)F_{\mathrm{B}}(T)
\label{eq:rate-overall-ratio}
\end{equation}
Setting derivative of Equation \eqref{eq:rate-overall} with respect to
temperature to be zero at $T_{\mathrm{min}}$, one can obtain working
equation of temperature. Solving it with $\lambda$ as parameter,
$T_{\mathrm{min}}$ is computed. However, this woking equation is only
able to predict one trivial solution of $T_{\mathrm{min}}=0\;\mathrm{K}$
because of expression
\begin{equation}
\frac{\partial}{\partial T}J_{\lambda}(T)\propto\frac{\partial}{\partial T}
\Big(M(\lambda,T)F_{\mathrm{B}}(T)\Big)=
F_{\mathrm{B}}\frac{\partial M}{\partial T}+MF_{\mathrm{B}}
\frac{\Delta E}{k_{\mathrm{B}}T^2}=MF_{\mathrm{B}}
\left(\frac{b\lambda^4}{\gamma T^2}M\exp\left(\frac{b}{\lambda T}\right)
+\frac{\Delta E}{k_{\mathrm{B}}T^2}\right)>0.
\label{eq:rate-overall-ratio-11}
\end{equation}
Substituting Equation \eqref{eq:radiation-dis-bb} into Equation
\eqref{eq:rate-overall-ratio-11}, one has limitation of
$\partial J_{\lambda}(T)/\partial T\to0$ at absolute zero, in agreement
with the fact that rate constant of any reaction should be zero at
absolute zero. This is not surprising because $\partial J_{\lambda}(T)/\partial T=0$
means the impossible dependence of $T_{\mathrm{min}}$ on $\lambda$.
On the other hand, turning to the case at high temperature, where the
endothermic processes open due to existence of HO$_2$ ($\tilde{A}\;{}^2A'$),
the overall rate constant should be changed at the temperature where
the endothermic process with the lowest barrier ({\it i.e.}, Reaction
\eqref{eq:o2-wat-XAcX} with $0.107$ eV) opens. The barrier of $0.107$
eV is related with temperature of 1242 K which is also able to open
the HO$_2$ transition $\tilde{A}\;{}^2A'\leftarrow\tilde{X}\;{}^2A''$
as shown by Figure \ref{fig:blackbody}. In this case, the endothermic
processes play a role in the model above $1242$ K. Therefore,
at the temperature range from $900$ K until $1242$ K, the HO$_2$
transition $\tilde{A}\;{}^2A'\leftarrow\tilde{X}\;{}^2A''$ occurs but
the endothermic processes are still closed making the overall rate
constant decrease. This indicates existence of the well in the Arrhenius
curve at the temperature range from $900$ K to $1242$ K. 
These results are in good agreement with the experimental
measurements within the temperature range of $900$ K to $1250$ K
(see Figure \ref{fig:expt-arr-plot}).

\subsection{Recombination and Nonadiabatic Pathways\label{sec:quan-dyn}}

Having interpretation the temperature-depedence of rate constant, now
we turn to recombination and nonadiabatic processes which are supplementary
pathways of the above model. First of all, the above black-body model
has predicted a factor of $\sim2.7$ for the ratio
$k(T=300\;\mathrm{K})/k(T=1100\;\mathrm{K})$ which approaches to the
experimental upper-limitation of $\sim3.3$. This factor can be further
reduced if one notes roles of the dissociation and recombination of HO$_2$
in obtaining HO$_2$ ($\tilde{A}\;{}^2A'$). As shown in Tables \ref{tab:mrci-details}
and \ref{tab:net-hox-kin}, the HO$_2$ dissociation is primary or secondary
because of rather small energy difference. Figure \ref{fig:corr-diag}(a)
indicates that the reactants at the ground state, OH ($\tilde{X}\;{}^2\Pi$)
+ HO$_2$ ($\tilde{X}\;{}^2A''$), can form O$_2$ ($\tilde{a}\;{}^1\Delta_g$)
which can be further react with the H (${}^2S_u$) atom to finally obtain
HO$_2$ ($\tilde{A}\;{}^2A'$), namely reverse Reaction \eqref{eq:diss-of-ho2-Xa},
\begin{equation}
\mathrm{H}\;({}^2S_u)+\mathrm{O}_2\;(\tilde{a}\;{}^1\Delta_g)+\mathrm{M}\to
\mathrm{HO}_2\;(\tilde{A}\;{}^2A')+\mathrm{M},\quad\Delta E=-2.361\;\mathrm{eV},
\label{eq:recombina-ho2-A}
\end{equation}
where an inactive special M is required, such as one water molecule.
This pathway is possible due to the fact that both Reactions \eqref{eq:o2-wat-XXaX}
and \eqref{eq:recombina-ho2-A} are exothermic without barriers. Thus,
the channels starting from HO$_2$ ($\tilde{A}\;{}^2A'$) open at low
temperature implying that $k(T=300\;\mathrm{K})$ and hence the ratio
$k(T=300\;\mathrm{K})/k(T=1100\;\mathrm{K})$ may decrease approaching
to its experimental lower-limitation of $1.23$. However, the process
for obtaining HO$_2$ ($\tilde{A}\;{}^2A'$) with Reaction \eqref{eq:recombina-ho2-A}
requires additional H (${}^2S_u$) in the mechanism proposed in
Sections \ref{sec:dynamics} and \ref{sec:kinetics-exp}. Since H
(${}^2S_u$) is richly distributed in the combustion systems and
can be obtained from dissociations of HO$_2$
(see Reactions \eqref{eq:diss-of-ho2-XX} to \eqref{eq:diss-of-ho2-aX}),
this recombination process of Reaction \eqref{eq:recombina-ho2-A}
probably exists.

Second, as illustrated in Figure \ref{fig:corr-diag}(a) two crossing points
of $3\;{}^1A/2\;{}^3A$ and $4\;{}^1A/2\;{}^3A$ exist in the present model.
These crossing points indicate the $3\;{}^1A/2\;{}^3A$
and $4\;{}^1A/2\;{}^3A$ seams which might play an important role in
nonadiabatic dynamics of Reaction \eqref{eq:o2-wat} through spin-orbit
coupling (SOC). The present geometry optimizations at the MCSCF level on
the seams predict MECPs' geometries and the SOC values. Further, the
energy values are computed at the MRCI//MCSCF level where the energy
difference between two crossing states is approximately equal to or
smaller than $2.0$ kcal/mol (that is chemical accuracy). Due to rather
larger SOC values, nonadiabatic processes of Reaction \eqref{eq:o2-wat}
are allowable and can be given by
\begin{align}
\mathrm{OH}\;(\tilde{X}\;{}^2\Pi)+\mathrm{HO}_2\;(\tilde{X}\;{}^2A'')
&\to\mathrm{HO}_x\;(2\;{}^3A)\to3\;{}^1A/2\;{}^3A\;\mathrm{MECP}
\allowdisplaybreaks[4]  \nonumber \\
&\to\mathrm{HO}_x\;(3\;{}^1A)\to\mathrm{O}_2\;(\tilde{b}\;{}^1\Sigma_g^+)
+\mathrm{H}_2\mathrm{O}\;(\tilde{X}\;{}^1A_1),\quad\Delta E=8.587\;\mathrm{eV},
\label{eq:o2-wat-XXbX-nonad}  \allowdisplaybreaks[4]  \\
\mathrm{OH}\;(\tilde{X}\;{}^2\Pi)+\mathrm{HO}_2\;(\tilde{X}\;{}^2A'')
&\to\mathrm{HO}_x\;(2\;{}^3A)\to4\;{}^1A/2\;{}^3A\;\mathrm{MECP}
\allowdisplaybreaks[4]  \nonumber \\
&\to\mathrm{HO}_x\;(4\;{}^1A)\to\mathrm{O}_2\;(\tilde{c}\;{}^1\Sigma_u^-)
+\mathrm{H}_2\mathrm{O}\;(\tilde{X}\;{}^1A_1),\quad\Delta E=9.273\;\mathrm{eV},
\label{eq:o2-wat-XXcX-nonad}
\end{align}
where the SOC values of the $3\;{}^1A/2\;{}^3A$ and $4\;{}^1A/2\;{}^3A$
MECPs are $\sim400$ cm$^{-1}$ and $\sim620$ cm$^{-1}$. By Reactions
\eqref{eq:o2-wat-XXbX-nonad} and \eqref{eq:o2-wat-XXcX-nonad} starting
from OH ($\tilde{X}\;{}^2\Pi$) + HO$_2$ ($\tilde{X}\;{}^2A''$), it is
possible to obtain O$_2$ ($\tilde{b}\;{}^1\Sigma_g^+$) and O$_2$ ($\tilde{c}\;{}^1\Sigma_u^-$)
which are direct products from OH ($\tilde{X}\;{}^2\Pi$) + HO$_2$ ($\tilde{A}\;{}^2A'$).
However, as given in Table \ref{tab:net-hox-kin}, the nonadiabatic processes
of Reactions \eqref{eq:o2-wat-XXbX-nonad} and \eqref{eq:o2-wat-XXcX-nonad}
occur with rather low probability and need to be further inspected by
experiment due to the following two points. On the one hand, Reactions
\eqref{eq:o2-wat-XXbX-nonad} and \eqref{eq:o2-wat-XXcX-nonad} have very
high barriers at the MECPs which are energy minima on the seams. On the
other hand, these nonadiabatic processes require transitions between
singlet and triplet states via SOC making Reactions \eqref{eq:o2-wat-XXbX-nonad}
and \eqref{eq:o2-wat-XXcX-nonad} essentially forbidden. Despite of this,
Reactions \eqref{eq:o2-wat-XXbX-nonad} and \eqref{eq:o2-wat-XXcX-nonad}
play a role in reduce $k(T=300\;\mathrm{K})/k(T=1100\;\mathrm{K})$ because
these processes can open at room temperature. In previously reported
nonadiabatic dynamics \cite{son24:597}, the first singlet and triplet
states are involved in calculations, called S$_0$ and T$_1$. According
to the present MRCI calculations, the S$_0$ and T$_1$ states are
HO$_x$ ($1\;{}^1A$) and HO$_x$ ($1\;{}^3A$), respectively. Figure
\ref{fig:corr-diag}(a) indicates that reactions at the S$_0$ and
T$_1$ states represent Reaction \eqref{eq:o2-wat-XXaX} and
\eqref{eq:o2-wat-XXXX}, respectively, in which SOC may play unimportant
role. This is in consistent with previous dynamics calculations \cite{son24:597}
based on $2\times2$ Hamiltonian matrix. However, further quantum dynamics
calculations must be performed to quantitatively confirm role of the
present nonadiabatic processes in Reactions \eqref{eq:o2-wat-XXbX-nonad} and
\eqref{eq:o2-wat-XXcX-nonad}. This is planning in the future-coming
ML-MCTDH calculations with a new $8\times8$ Hamiltonian matrix.

\subsection{Discussions and Perspectives\label{sec:disscu}}

Next, we should further discuss on the present model for
explaining the Arrhenius curve in Figure \ref{fig:expt-arr-plot}.
First of all, comparing with previously reported studies
\cite{liu19:12667,son20:134309,liu20:3331,son24:597} on reaction
dynamics of Reaction \eqref{eq:o2-wat}, the present model is proposed
on the basis of electronic-structure calculations without the aid of
molecular dynamics or quantum dynamics simulations.
Although it only predicts from the viewpoints of electronic
energy, quantitative predictions on the rate constant are given if
our previously reported RPMD rate constants are introduced. As given
in Section \ref{sec:kinetics-exp}, the
present model generally requires the following two
points. On the one hand, pathway at the ground-state, such as Reaction
\eqref{eq:o2-wat-XXXX}, should be strongly exothermic,
which indicates weak temperature-dependence or temperature-independence
of the rate constant at low temperature (say the room
temperature). A kinetics model \cite{son20:134309} to interpret this
feature of Reaction \eqref{eq:o2-wat} was proposed by RPMD simulations
based on the GPR PES at the ground state. However, pathways at excited
states should be weakly endothermic. These pathways may open and play
an important role at high temperature, as shown by the present model.
To this end, on the other hand, one of reactant fragments should be
easily excited to its excited electronic-state at higher temperatures,
say $\sim10^3$ K. Since characteristic temperature for electronic
motion is typical $\sim10^4$ K, electronic excitation at $\sim10^3$ K
requires some special mechanism to produce photon. The
present model assumes that photon wavelength in the system satisfies
the black-body distribution as function of temperature. According to
Planck radiation distribution, the present model allows the
transition HO$_2$ ($\tilde{A}\;{}^2A'$) $\leftarrow$ HO$_2$
($\tilde{X}\;{}^2A''$) to occur at $\sim1100$ K. There might exist
other mechanism in consistent with system characteristic, such as
the recombination process in Reaction \eqref{eq:recombina-ho2-A}.

Second, the present model can be extended to other similar
reactions. Some probable examples are collected in the Supporting
Information. For instance, the H + O$_2$ + M $\to$ HO$_2$ + M reaction,
the recombination process of the present model, {\it i.e.} Reaction
\eqref{eq:recombina-ho2-A}, clearly has a well in its Arrhenius curve
\cite{bau05:757} at $\sim1200$ K if M is Ar or H$_2$O. Obviously, if
the system meets the above two requirements, the present model can be
adopted. This implies that excited electronic-states play an important
role in dynamics and kinetics, even though the system is dark, as long as there
exists a mechanism to produce the exicted states. In this context,
before studying reaction dynamics and mechanism through electronic-structure
calculations, an approximately complete Hilbert subspace for electronic
motions must be determined for a special system. In this work, Reaction
\eqref{eq:o2-wat}, the 8D Hilbert subspace is determined for the present
model implying that previous space containing only S$_0$ and
T$_1$ is incomplete to predict reasonable dynamics feature.
Although the present model can explain many features of the
Arrhenius curve, state/mode-specific quantum dynamics must be performed
for further quantitative dynamics results. Based on the present MRCI
calculations in conjunction with previous studies \cite{son20:134309,son24:597}
on Reaction \eqref{eq:o2-wat}, construction of the $8\times8$ potential
matrix and theoretical developments on new 10D ML-MCTDH calculations
are planning in future work.
It should be finally mentioned that such new 10D ML-MCTDH calculations
are unable to contain the recombination process in Reaction \eqref{eq:recombina-ho2-A}
because the coordinates frame (see Figure \ref{fig:MLtree})
was designed only for the title reaction. Since Reaction \eqref{eq:recombina-ho2-A}
plays an important role in reducing the ratio
$k(T=300\;\mathrm{K})/k(T=1100\;\mathrm{K})$, the state/mode-specific
quantum dynamics might overestimate the rate constant.

Finally, noting that the present model employs a typical bottom-up
approach, we must discuss on the explanation for the same problem
suggested by Burke and co-workers \cite{bur13:547}, where a top-down
(or called up-bottom) approach was employed. In general, kinetics
mechanism of a complex reaction, say combustion, is comprised of
dozens to thousands of elementary reactions with rate constants
determined by a wide variety of sources. To integrate these data
in a self-consistent manner, Burke and co-workers \cite{bur13:547}
introduced a comprehensive framework that encompasses behaviors
spanning multiple chemically relevant scales, ranging from fundamental
molecular interactions to complex global combustion phenomena. Employing
such top-down approach, although the temperature dependence is
substantially less pronounced than previously suggested, Burke and
co-workers \cite{bur13:547} found a rate minimum near $1200$ K and
provided a quantitative rate constant in a manner consistent with
experimental data from the entire temperature range and transition-state
theory within their associated uncertainties. This top-down approach
\cite{bur13:547} and the present bottom-up approach represent two
distinct ways of understanding and solving rate temperature-dependence
of Reaction \eqref{eq:o2-wat}. The top-down approach \cite{bur13:547}
starts with a multi-scale model with specific restrictions and includes
a set of kinetics parameters with constrained uncertainties. These
parameters are connected by temperature-, pressure-, and bath-gas-dependent
rate constants, which have propagated uncertainties. The rate constants
are then related with combustion behavior, also with propagated uncertainties.
Direct incorporation yields more reliable extrapolation of limited data
to conditions outside the validation set. This is particularly helpful
for extrapolating to engine-relevant conditions where relatively limited
data are available. Unlike previous multi-scale top-down approach
\cite{bur13:547}, the present bottom-up model builds understanding of
the rate temperature-dependence of Reaction \eqref{eq:o2-wat} from the
electronic structure starting with fundamental state-to-state processes
and gradually assembling them into a larger picture. As given in Section
\ref{sec:theore-work}, we firstly optimize
geometries of fragments of the HO$_x$ system, then obtain correlation
diagram, and finally explain the rate temperature-dependence by the
present model with black body. While the multi-scale top-down approach
\cite{bur13:547} emphasizes abstraction, the present approach focuses on
detailed state-to-state processes and thus incremental construction.
Therefore, the multi-scale top-down approach \cite{bur13:547} provides
a perspective on role of Reaction \eqref{eq:o2-wat} in a special combustion
mechanism with given conditions, while the present bottom-up approach
ensures a more foundational understanding. One might combine these ways
to tackle the present problem, leveraging the strengths of each to
achieve a more comprehensive understanding.

\section{Conclusions\label{sec:con}}

In this work, a model with multiple electronic states for the OH + HO$_2$ $\to$ O$_2$ + H$_2$O
reaction is proposed on the basis of extensive MRCI/aug-cc-pVTZ calculations.
Its main purpose is to explain the deep and unusually narrow well in
the Arrhenius curve near $1100\:\mathrm{K}$ together with slightly
negative temperature-dependence at $200\:\mathrm{K}\sim500\:\mathrm{K}$,
as shown in Figure \ref{fig:expt-arr-plot}. The MRCI optimizations for
geometries of fragments in the HO$_x$ system are perfomed to obtain
state-to-state scattering processes starting from the first two states of
HO$_2$, namely $\tilde{X}\;{}^2A''$ and $\tilde{A}\;{}^2A'$. These
processes compose an eight-dimensional Hilbert subspace which is nearly
independent and decoupled from higher-lying electronic-states. As
shown by Figure \ref{fig:corr-diag}(a), three of four processes with
HO$_2$ ($\tilde{X}\;{}^2A''$) are exothermic, while those with HO$_2$
($\tilde{A}\;{}^2A'$) have three endothermic channels with the smallest
barrier of $0.107$ eV. At low temperature below $500$ K, processes with
excited HO$_2$ ($\tilde{A}\;{}^2A'$) are closed and its exothermic
feature makes rate constant temperature-independence. A black-body
scheme is introduced to allow excitation of HO$_2$ at $>900$ K making
endothermic processes with barrier above 0.107 eV occur and rate
constant accordingly reduced by a factor of $0.3711$ at $1100$ K.
This factor demonstrates close agreement with the
experimentally observed range of $0.30\sim0.76$. At
high temperature above $1242$ K, which is associated with the barrier
of 0.107 eV, the endothermic processes open and the overall rate
constant depends on the temperature as Boltzmann factor leading to the
well in the Arrhenius curve at the range from $900$ K to $1242$ K. This
range of temperature is very close to experimental measurements of
$900\;\mathrm{K}\sim1250\;\mathrm{K}$. In addition, we also propose
recombination and nonadiabatic processes as another pathways to
obtain HO$_2$ ($\tilde{A}\;{}^2A'$) and form O$_2$ at higher-lying
states, $\tilde{b}\;{}^1\Sigma_g^+$ and $\tilde{c}\;{}^1\Sigma_u^-$,
respectively. Finally, discussions on the role of multi-electronic-state
model in combustion chemistry are given leading to a general model to
explain similar phenomena. Moreover, we compare the present approach
with previously reported multi-scale top-down approach to show difference
between these two distinct ways on chemical dynamics.

\section*{Supplementary Material}

See Supporting Information at http://dx.doi.org/XXX for
(1) details of previously reported measurements shown in Figure
\ref{fig:expt-arr-plot} and
(2) details of the present MRCI results on state-to-state scattering
processes shown in Figure \ref{fig:corr-diag}.

\section*{Interest Statement}

The authors declare that they have no known competing financial interests
or personal relationships that could have appeared to influence the work
reported in this paper.

\section*{Data Availability}

All data have been reported in this work. The data supporting this article
have been included as part of the Supplementary Information.

\section*{Acknowledgements}

The financial supports of National Natural Science Foundation of China
(Grant No. 22273074) and Fundamental Research Funds for the Central
Universities (Grant No. 2025JGZY34) are gratefully acknowledged. We
are grateful to the anonymous reviewers for their thoughtful suggestions.


\clearpage
\begin{sidewaystable}
\caption{
Numerical details of the present MRCI calculations, together with
dissociation limitations of each fragment. The first column gives
fragments of the HO$_x$ system. The second and third columns give
symmetries of each fragment and those in MRCI calculations, respectively.
The fourth column gives electronic wave function of each fragment at
the ground state, where only the occupancies of the valence orbitals
are given. The fifth column gives reference coefficient as characteristics
of the electronic wave function for the fragment at the ground state.
The sixth column gives the active space in the present MRCI calculations
in the form of ($n$o,$m$e) meaning $n$ active orbitals occupied by $m$
active electrons. The seventh and eighth columns give electronic state of
the dissociation limit and the relative energy values (in eV) to the
fragment itself. The rightmost column gives reactive characteristic
of the HO$_x$ system according to energy profile of the processes. 
}%
 \begin{tabular}{llclclrlllllrrrrr}
  \hline
Frag. &~~& \multicolumn{3}{c}{Symmetry} &~~& \multicolumn{5}{c}{Wave Function} 
&~~& \multicolumn{3}{c}{Dissociation Limit} &~~& Charact.
\\ \cline{3-5}\cline{7-11}\cline{13-15}
&& Group &~~& Sub. && Ground State &~~& Coeff. &~~& Act. Sp. && State &~~& Energy && \\
\hline
H$_2$  && $D_{\infty\mathrm{h}}$ && $D_{2\mathrm{h}}$ && $(1\sigma_g)^2(1\sigma_u)^0$ 
&& $0.990$ && (2o,2e)
&& H (${}^2S_u$) + H (${}^2S_u$) && $4.707$ && Thirdhand
\footnote{The process with the relative energy in the range from 1.5 eV
to 4.5 eV or form 4.5 eV to 10.0 eV is called secondary or thirdhand,
respectively.\label{foot:second-third}} \\
O$_2$  && $D_{\infty\mathrm{h}}$ && $D_{2\mathrm{h}}$ &&
$KK(2\sigma_g)^2(2\sigma_u)^2(3\sigma_g)^2(1\pi_u)^4(1\pi_g)^2(3\sigma_u)^0$ && $0.991$ && (12o,8e)
&& O (${}^3P_g$) + O (${}^3P_g$) && $27.998\sim36.103$
\footnote{Range of dissociation energy of O$_2$ below the $\tilde{d}\;{}^1\Pi_g$ state.
All of these states dissociate to the O (${}^3P_g$) atom.}  && Least
\footnote{The processes with the relative energy above 10.0 eV are called least.
\label{foot:least}} \\
OH     && $C_{\infty\mathrm{v}}$ && $C_{2\mathrm{v}}$ && $K(2\sigma)^2(3\sigma)^2(1\pi)^3(4\sigma)^0$
&& $0.999$ && (5o,7e) && H (${}^2S_u$) + O (${}^3P_g$) && $16.095$ and $20.227$
\footnote{Dissociation energies of OH ($\tilde{X}\;{}^2\Pi$) and OH ($\tilde{A}\;{}^2\Sigma^+$)
dissociating to H (${}^2S_u$) + O (${}^3P_g$).}  && Least \textsuperscript{\ref{foot:least}} \\
HO$_2$ && $C_\mathrm{s}$          && $C_{\mathrm{s}}$ && $KK(3a')^2(4a')^2(5a')^2(1a'')^2$ && 0.995
&& (9o,13e) && H (${}^2S_u$) + O$_2$ ($\tilde{X}$/$\tilde{a}$/$\tilde{A}'$)
&& $0.179\sim2.361$ 
\footnote{Range of dissociation energy of HO$_2$ at the $\tilde{X}\;{}^2A''$,
$\tilde{A}\;{}^2A'$, and $\tilde{a}\;{}^4A''$ dissociating to O$_2$ at
$\tilde{X}\;{}^3\Sigma_g^-$, $\tilde{a}\;{}^1\Delta_g$, and $\tilde{X}\;{}^3\Sigma_g^-$, 
respectively.}   && Primary
\footnote{The process with the smallest relative energy (below 1.5 eV) is called
primary.\label{foot:primary}} \\
&&  &&  && $(6a')^2(7a')^2(2a'')^1(8a')^0(9a')^0$ &&  &&   &&  &&  && 
or Secondary \textsuperscript{\ref{foot:second-third}} \\
H$_2$O && $C_{2\mathrm{v}}$ && $C_{2\mathrm{v}}$ &&  $K(2a_1)^2(1b_2)^2(3a_1)^2(1b_1)^2(4a_1)^0(2b_2)^0$
&& $0.998$ && (6o,8e) && H (${}^2S_u$) + OH ($\tilde{X}\;{}^2\Pi$) && $5.259$ && Thirdhand
\textsuperscript{\ref{foot:second-third}} \\
O$_3$  && $C_{2\mathrm{v}}$       && $C_{2\mathrm{v}}$ &&
$KKK(3a_1)^2(2b_2)^2(4a_1)^2(5a_1)^2(1b_1)^2$ && $0.992$ && (12o,18e) && 
O (${}^3P_g$) + O$_2$ ($\tilde{X}$/$\tilde{a}$) && $14.530\sim16.120$
\footnote{Range of dissociation energy of O$_3$ below the
$\tilde{B}\;{}^1B_1$ state dissociating to O (${}^3P_g$) and O$_2$ at
the $\tilde{X}\;{}^3\Sigma_g^-$ and $\tilde{a}\;{}^1\Delta_g$ states.} && Least
\textsuperscript{\ref{foot:least}} \\
&&  &&  && $(3b_2)^2(4b_2)^2(6a_1)^2(1a_2)^2(2b_1)^0(7a_1)^0(5b_2)^0$ && &&  &&  &&  && \\
HO$_x$ && $C_1$ && $C_1$ && $KKK(4a)^2(5a)^2(6a)^2(7a)^2(8a)^2(9a)^2(10a)^2$
&& $0.900$ && (12o,16e) && O$_2$ + H$_2$O && $-3.035\sim1.437$
\footnote{See also the Supporting Information.\label{note:see-si}} && Primary
\textsuperscript{\ref{foot:primary}} \\
&&  &&  && $(11a)^2(12a)^2(13a)^2(14a)^0(15a)^0(16a)^0(17a)^0$ &&  && && H$_2$ + O$_3$ && $1.623\sim4.231$
\textsuperscript{\ref{note:see-si}}  && Secondary \textsuperscript{\ref{foot:second-third}} \\ 
\hline
 \end{tabular}
   \label{tab:mrci-details}
    \end{sidewaystable}
  
\clearpage
\begin{table}
\caption{
State-to-state processes for primary processes (upper panel) and other
cases (lower panel) in the present model of the OH + HO$_2$
$\to$ O$_2$ + H$_2$O reaction, together with energy profile (in eV)
and corresponding wavelength (in nm) as well as associated black-body
temperature (in K). The OH + HO$_2$ system only has identical symmetry
operation and thus belongs to the C$_1$ group. According to symmetry
of each channel, one can obtain the correlation diagram as shown in
Figure \ref{fig:corr-diag} by which eigth states are chosen to form
the present Hilbert subspace. The second column gives state-to-state
processes of the present Hilbert subspace (upper pannel), together
with secondary channel of H$_2$ + O$_3$ and dissociation channel of
HO$_2$ (lower panel). The third, fourth, and fifth columns give the
energy values (in eV) relative to O$_2$ ($\tilde{X}\;{}^3\Sigma_g^-$)
+ H$_2$O ($\tilde{X}\;{}^1A_1$), including energy values of initial
and final states and energy difference $\Delta E$ between them. The
sixth and seventh columns give energy values (in eV) relative to the
reactant channel OH ($\tilde{X}\;{}^2\Pi$) + HO$_2$ ($\tilde{X}\;{}^2A''$)
and corresponding wavelength $\lambda$ (in nm), respectively. The eighth
columns gives the black-body temperature $T_{\mathrm{max}}$ at which
Planck radiation distribution has the maximum value at associated wavelength
$\lambda$ given in the seventh column. Wien displacement law predicts
that $T_{\mathrm{max}}=2.8977721\times10^{6}\;\mathrm{K/nm}/\lambda$.
The rightmost column give remarks for each process, including thermodynamics
characteristic, oscillator strength $f$, SOC values (in cm$^{-1}$),
and Boltzmann distribution $\exp(-\Delta E/k_{\mathrm{B}}T)$ associated
with each process at 1100 K as well as order-of-magnitude difference
(in parentheses) between it and those at 300 K.
}
\label{tab:net-hox-kin}
\end{table}
\clearpage
\begin{sidewaystable}
\begin{tabular}{ccllrlrlrlrlrlrrr}
\hline
No. &~~& Reaction &~~& \multicolumn{5}{c}{Energy} &~~& \multicolumn{7}{c}{Model} \\ 
\cline{5-9}\cline{11-17}
    &&  && Reactants && Products && Difference
\footnote{It is either energy difference between products and
reactants or potential barrier during reaction process.}
    && Energy && $\lambda$ && $T_{\mathrm{max}}$ && Remark \\
\hline
\multicolumn{17}{l}{{\it Primary Processes}}  \\
1&&OH ($\tilde{X}\;{}^2\Pi$) + HO$_2$ ($\tilde{X}\;{}^2A''$) $\to$
O$_2$ ($\tilde{X}\;{}^3\Sigma_g^-$) + H$_2$O ($\tilde{X}\;{}^1A_1$)
&& $3.035$ && $0.000$ && $-3.035$ && --- && --- && --- && exothermic \\
2&&OH ($\tilde{X}\;{}^2\Pi$) + HO$_2$ ($\tilde{X}\;{}^2A''$) $\to$
O$_2$ ($\tilde{a}\;{}^1\Delta_g$) + H$_2$O ($\tilde{X}\;{}^1A_1$)
&& $3.035$ && $0.985$ && $-2.050$ && --- && --- && --- && exothermic \\
3&&OH ($\tilde{X}\;{}^2\Pi$) + HO$_2$ ($\tilde{A}\;{}^2A'$) $\to$
O$_2$ ($\tilde{b}\;{}^1\Sigma_g^+$) + H$_2$O ($\tilde{X}\;{}^1A_1$)
&& $3.883$ && $1.115$ && $-2.768$ && --- && --- && --- && exothermic \\
4&& HO$_2$ ($\tilde{A}\;{}^2A'$) $\leftarrow$ HO$_2$ ($\tilde{X}\;{}^2A''$)
&& $3.035$ && $3.883$ && $0.848$ && $0.848$ && $1462.26$ && $1981.7$ && $f=0.987$ \\
5&&OH ($\tilde{X}\;{}^2\Pi$) + HO$_2$ ($\tilde{A}\;{}^2A'$) $\to$
O$_2$ ($\tilde{c}\;{}^1\Sigma_u^-$) + H$_2$O ($\tilde{X}\;{}^1A_1$)
&& $3.883$ && $3.990$ && $0.107$ && $0.955$ && $1298.27$ && $2232.0$ 
&& $3.234\times10^{-1}$ $(-1)$ \\
6&&OH ($\tilde{X}\;{}^2\Pi$) + HO$_2$ ($\tilde{A}\;{}^2A'$) $\to$
O$_2$ ($\tilde{A}'\;{}^3\Delta_u$) + H$_2$O ($\tilde{X}\;{}^1A_1$)
&& $3.883$ && $4.205$ && $0.322$ && $1.170$ && $1059.70$ && $2734.5$ 
&& $3.347\times10^{-2}$ $(-4)$ \\
7&&OH ($\tilde{X}\;{}^2\Pi$) + HO$_2$ ($\tilde{X}\;{}^2A''$) $\to$
O$_2$ ($\tilde{A}'\;{}^3\Delta_u$) + H$_2$O ($\tilde{X}\;{}^1A_1$)
&& $3.035$ && $4.205$ && $1.170$ && $1.170$ && $1059.70$ && $2734.5$ 
&& $4.360\times10^{-6}$ $(-14)$ \\
8&&OH ($\tilde{X}\;{}^2\Pi$) + HO$_2$ ($\tilde{A}\;{}^2A'$) $\to$
O$_2$ ($\tilde{A}\;{}^3\Sigma_u^+$) + H$_2$O ($\tilde{X}\;{}^1A_1$)
&& $3.883$ && $4.286$ && $0.403$ && $1.251$ && $991.08$ && $2923.9$
&& $1.424\times10^{-2}$ $(-5)$ \\
\hline
\multicolumn{17}{l}{{\it Secondary Processes}}  \\
9&&OH ($\tilde{X}\;{}^2\Pi$) + HO$_2$ ($\tilde{X}\;{}^2A''$) $\to$
H$_2$ ($\tilde{X}\;{}^1\Sigma_g^+$) + O$_3$ ($\tilde{X}\;{}^1A_1$)
&& $3.035$ && $4.658$ && $1.623$ && $1.623$ && $763.92$ && $3793.3$ 
&& $3.665\times10^{-8}$ $(-20)$ \\
10&&HO$_2$ ($\tilde{X}\;{}^2A''$) $\to$ H (${}^2S_u$) + O$_2$
($\tilde{X}\;{}^3\Sigma_g^-$)
&& --- && --- && $2.224$ && $2.224$ && $557.48$ && $5198.0$ && $6.464\times10^{-11}$ $(-27)$ \\
11&&HO$_2$ ($\tilde{A}\;{}^2A'$) $\to$ H (${}^2S_u$) + O$_2$
($\tilde{a}\;{}^1\Delta_g$)
&& --- && --- && $2.361$ && $2.361$ && $525.13$ && $5518.2$ && $1.523\times10^{-11}$ $(-27)$ \\
12&& H (${}^2S_u$) + O$_2$ ($\tilde{a}\;{}^1\Delta_g$) + M $\to$
HO$_2$ ($\tilde{A}\;{}^2A'$) + M
\footnote{The inactive molecule M might be Ar or H$_2$O.}
&& --- && --- && $-2.361$ && --- && --- && --- && exothermic \\
13&& $\mathrm{OH}\;(\tilde{X}\;{}^2\Pi)+\mathrm{HO}_2\;(\tilde{X}\;{}^2A'')\to$
&& $3.035$ && $1.115$ && $11.622$
\footnote{The MRCI energy difference between the two crossing states
is $0.087$ eV ($\sim2.0$ kcal/mol) at the MCSCF optimized
$3\;{}^1A/2\;{}^3A$ MECP geometry.\label{foot:3s-2t}}
&& $8.587$ \textsuperscript{\ref{foot:3s-2t}}
&& $144.39$ && $20069$ && barrier at MECP \\
&& $\mathrm{HO}_x\;(2\;{}^3A)\to3\;{}^1A/2\;{}^3A\;\mathrm{MECP}$
\footnote{Geometries of the MECPs are optimized at the MCSCF level, while relative
energies of the MECPs are computed at the MRCI//MCSCF level.\label{foot:mecps}}
$\to\mathrm{HO}_x\;(3\;{}^1A)$
&&  &&  &&  &&  &&  &&  && $\sim0.0$ \\
&& $\to\mathrm{O}_2\;(\tilde{b}\;{}^1\Sigma_g^+)
+\mathrm{H}_2\mathrm{O}\;(\tilde{X}\;{}^1A_1)$
&&  &&  &&  &&  &&  &&  && $\sim400$ cm$^{-1}$ \\
14&& $\mathrm{OH}\;(\tilde{X}\;{}^2\Pi)+\mathrm{HO}_2\;(\tilde{X}\;{}^2A'')\to$
&& $3.035$ && $3.990$ && $12.308$ 
\footnote{The MRCI energy difference between the two crossing states
is $0.039$ eV ($\sim0.9$ kcal/mol) at the MCSCF optimized
$4\;{}^1A/2\;{}^3A$ MECP geometry.\label{foot:4s-2t}}
&& $9.273$ \textsuperscript{\ref{foot:4s-2t}} && $133.70$ && $21674$ && barrier at MECP \\
&& $\mathrm{HO}_x\;(2\;{}^3A)\to4\;{}^1A/2\;{}^3A\;\mathrm{MECP}$
\textsuperscript{\ref{foot:mecps}}
$\to\mathrm{HO}_x\;(4\;{}^1A)$
&&  &&  &&  &&  &&  &&  && $\sim0.0$ \\
&& $\to\mathrm{O}_2\;(\tilde{c}\;{}^1\Sigma_u^-)
+\mathrm{H}_2\mathrm{O}\;(\tilde{X}\;{}^1A_1)$
&&  &&  &&  &&  &&  &&  && $\sim620$ cm$^{-1}$ \\
\hline
\end{tabular}
\end{sidewaystable}

\clearpage
 \section*{Figure Captions}
 
\figcaption{fig:expt-arr-plot}{%
Arrhenius plot of the rate constants of the OH + HO$_2$ $\to$ O$_2$ +
H$_2$O reaction, where the ordinate axis gives the temperature (in K)
while the abscissa axis gives logarithmic value of the rate constant
(in $\mathrm{cm}^2\cdot\mathrm{molecule}^{-1}\cdot\mathrm{s}^{-1}$).
The rate constants at low temperature range from $250\:\mathrm{K}$ to
$500\:\mathrm{K}$ have been well reviewed previously. Disregarding the
very low values, the remaining data at low temperature fall into two
groups of $\sim7\times10^{-11}\:\mathrm{cm}^2\cdot\mathrm{molecule}^{-1}\cdot\mathrm{s}^{-1}$
at low pressures and
$\sim1\times10^{-10}\:\mathrm{cm}^2\cdot\mathrm{molecule}^{-1}\cdot\mathrm{s}^{-1}$
at pressures close to one bar. This discrepancy was supposed to be due
to secondary chemistry of small amounts of H and O atoms in low-pressure,
rather than pressure dependence. At high temperature above $500\:\mathrm{K}$,
the rate constant initially declines in value as the temperature increases
reaching a minimum at $\sim1250\:\mathrm{K}$ and rises again at temperatures
beyond. This behavior was suggestive of formation and decomposition of
an intermediate complex, with the implication that the rate constant
is possibly pressure dependence. The preferred values at high temperatures
was suggeted to be
$8.3\times10^{-11}\:\mathrm{cm}^2\cdot\mathrm{molecule}^{-1}\cdot\mathrm{s}^{-1}$,
$1.0\times10^{-10}\:\mathrm{cm}^2\cdot\mathrm{molecule}^{-1}\cdot\mathrm{s}^{-1}$,
and $3.3\times10^{-11}\:\mathrm{cm}^2\cdot\mathrm{molecule}^{-1}\cdot\mathrm{s}^{-1}$.
further, a set of rather low values was observed by shock tube behind
reflected shock waves in conjunction with multi-species laser techniques,
as given by symbols $\Diamond$. All of citations and symbols shown in
the legend are given in the Supporting Information, where simple remarks
on these experiments are also given.
}%

\figcaption{fig:MLtree}{%
Hierarchical scheme in designing the coordinates sets. In this work,
since the total angular momentum of HO$_x$ is set to be zero and the
reaction occurs in the gas phase, the E$_2$ frame is equal to the SF
frame. Moreover, the OH and HO$_2$ fragments are described by
BF$_{\mathrm{OH}}$ and BF$_{\mathrm{HO}_2}$, respectively, while
relative frames BF$_{\mathrm{OH}}$/E$_2$ and BF$_{\mathrm{HO}_2}$/E$_2$
are defined to describe relative motions of OH and HO$_2$ with respect
with the E$_2$ frame, repsectively. Since the OH fragment is linear,
one of Euler angles in the BF$_{\mathrm{OH}}$/E$_2$ frame is arbitrary.
Further, due to the free motions of OH and HO$_2$, the relative frame
BF$_{\mathrm{OH}}$/BF$_{\mathrm{HO}_2}$ has to be designed to define
principal component of the reaction coordinate. We refer the reader to
the Supporting Information for more details on the present system and
its coordinates set.
}%

\figcaption{fig:corr-diag}{%
Correlation diagram of the low-lying state-to-state scattering processes
for (a) the OH + HO$_2$ $\to$ O$_2$ + H$_2$O reaction with the energy
of smaller than $7.000$ eV
and (b) the OH + HO$_2$ $\to$ H$_2$ + O$_3$ reaction with the energy
of smaller than $9.500$ eV.
The ordinate axis gives energy value in eV while the abscissa axis
qualitatively gives the reaction coordinate without showing any
probable intermediate and transition state. According to the
non-crossing rule of two states with the same symmetry, correlation
diagram connects the reactant and product fragments in different
electronic states. Since the symmetry must be maintained
throughout the reaction, it rationalizes the state-to-state pathways
to determine preferred ways or experimentally observed products for a
particular set of reactant fragments. The left and right panels give
the reactant fragment, OH + HO$_2$, and the product fragment, O$_2$
+ H$_2$O, respectively. We also give the electronic states and values
of relative energy (in eV) to the product in the ground state, O$_2$
($\tilde{X}\;{}^3\Sigma_g^-$) + H$_2$O ($\tilde{X}\;{}^1A_1$). The red
and blue dashed lines give correlation relations along the singlet and
triplet pathways, respectively, while the orange dashed line give those
along the quintet pathways.
}%

\begin{figure}
  \centering
   \includegraphics[width=18cm]{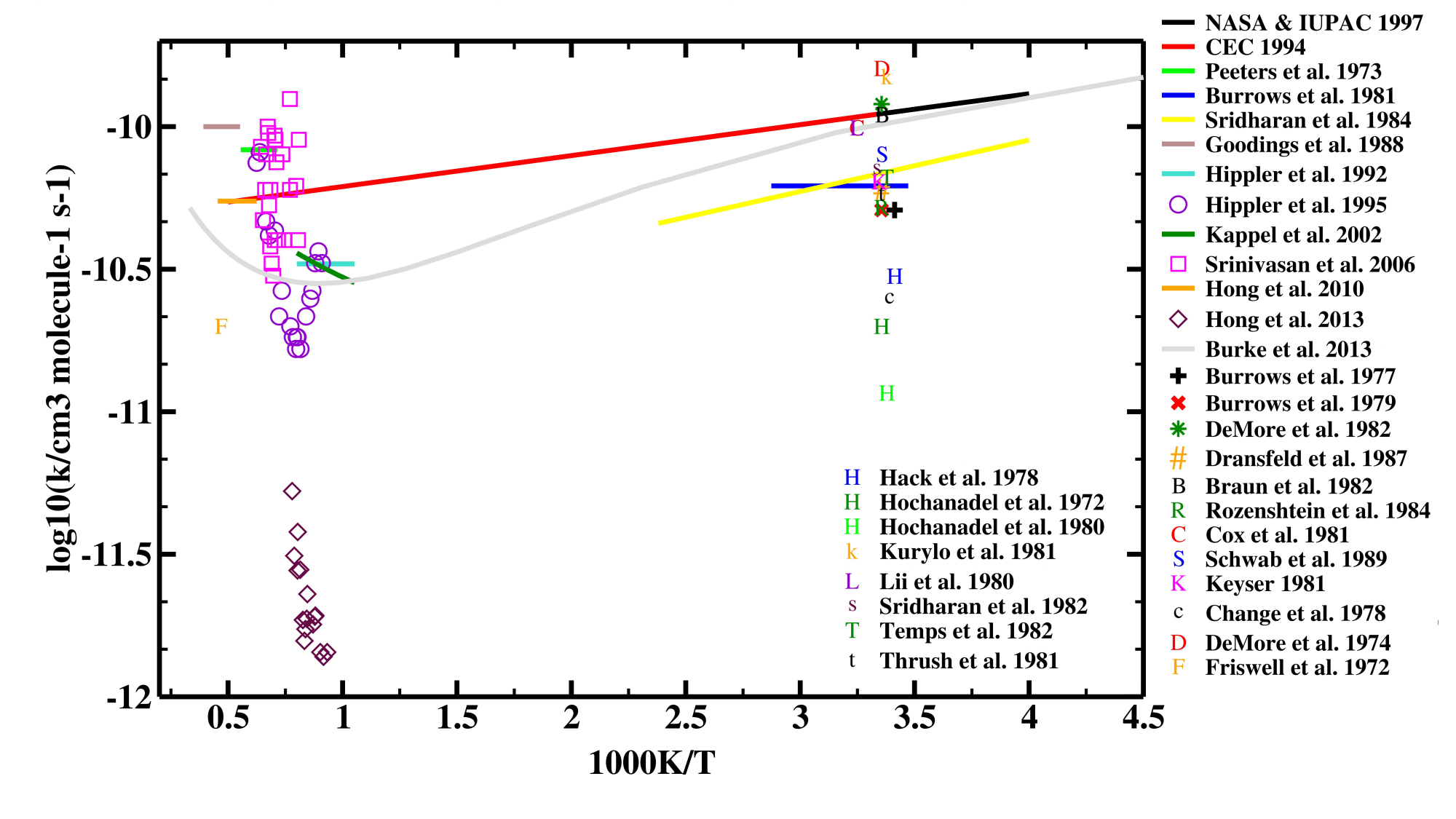}
    \caption{\figfoot}
     \label{fig:expt-arr-plot}
      \end{figure}

\figcaption{fig:blackbody}{%
Radiation distribution computed by Planck radiation law as function
of wavelength (in nm) and temperature (in K), where the combustion or
atmosphere system is supposed to be a black body and the energy zero
point is set to be eigen-energy of OH ($\tilde{X}\;{}^2\Pi$) + HO$_2$
($\tilde{X}\;{}^2A''$). For simple, the distributions are given as
functions of wavelength by coloered solid lines. The light green,
purple, green, blue, yellow, red, and brown lines represent distributions
at 300 K, 900 K, 1000 K, 1100 K, 1200 K, 1300 K, and 1500 K, respectively.
Moreover, gray dashed lines five wavelength values in understanding the
present model for interpretation of the Arrhenius curve (see
Figure \ref{fig:expt-arr-plot}). The wavelength values of 235.76 nm
and 557.48 nm are related with dissociation energies of HO$_2$
($\tilde{X}\;{}^2A''$) $\to$ H (${}^2S_u$) + O$_2$ ($\tilde{X}\;{}^3\Sigma_g^-$)
and H$_2$O ($\tilde{X}\;{}^1A_1$) $\to$ H (${}^2S_u$) + OH ($\tilde{X}\;{}^2\Pi$).
respectively. The wavelength values of 763.92 nm and 991.08 nm are
related with the first dissociation energy for H$_2$ + O$_3$ and the
largest value of the present Hilbert subspace, respectively. The
wavelength of 1462.24 nm is HO$_2$ excitation energy of transition
$\tilde{A}\;{}^2A'\leftarrow\tilde{X}\;{}^2A''$. The radiation
distribution as a function of wavelength has a maximum at point
$\lambda_{\mathrm{max}}$ which shifts with increasing temperature to
ever shorter wavelengths and depends on the temperature as Wien
displacement law. If $\lambda_{\mathrm{max}}=1426.26$ nm, the temperature
is then predicted to be 1981.7 K.
}%

\clearpage
      
\clearpage
  
\clearpage
      
\clearpage
                 

\clearpage

\begin{thebibliography}{10}
\ifnum\language=1 \def\biband{und}\else\def\biband{and}\fi

\bibitem{gar00}
{\rm J.~W.~C.~Gardiner}.
\newblock {\em Gas-Phase Combustion Chemistry}.
\newblock Springer, Heidelberg, 2000.

\bibitem{kon15:3755}
{\rm A.~A. Konnov}.
\newblock {On the role of excited species in hydrogen combustion}.
\newblock {\em Combust. Flame \bf 162\/} (2015), 3755--3772.

\bibitem{li21:105165}
{\rm G.~Li, M.~Fan, Y.~Lu,  \biband~ P.~Glarborg}.
\newblock {Kinetic modeling of carbon monoxide oxidation and water gas shift reaction in supercritical water}.
\newblock {\em J.\ Supercrit.\ Fluids \bf 171\/} (2021), 105165.

\bibitem{zha21:5799}
{\rm Y.~Zhang, J.~Fu, M.~Xie,  \biband~ J.~Liu}.
\newblock {Improvement of H$_2$/O$_2$ chemical kinetic mechanism for high pressure combustion}.
\newblock {\em Int.\ J.\ Hydrogen Energ. \bf 46\/} (2021), 5799--5811.

\bibitem{kli22:111975}
{\rm S.~J. Klippenstein, R.~Sivaramakrishnan, U.~Burke, K.~P. Somers, H.~J.
  Curran, L.~Cai, H.~Pitsch, M.~Pelucchi, T.~Faravelli,  \biband~ P.~Glarborg}.
\newblock {HO$_2$ + HO$_2$: High level theory and the role of singlet channels}. 
\newblock {\em Combust.\ Flame \bf 243\/} (2022), 111975.

\bibitem{li22:112093}
{\rm W.~Li, C.~Zou, H.~Yao, Q.~Lin, R.~Fu,  \biband~ J.~Luo}.
\newblock {An optimized kinetic model for H$_2$/CO combustion in CO$_2$ diluent at elevated pressures}.
\newblock {\em Combust.\ Flame \bf 241\/} (2022), 112093.

\bibitem{fer22:619}
{\rm A.~F. {de Ferreira Miranda}, G.~F. Bauerfeldt,  \biband~ L.~Baptista}.
\newblock {Numerical simulation of the gas-phase thermal decomposition and the detonation of H$_2$O$_2$/H$_2$O mixtures}.
\newblock {\em Reac.\ Kinet.\ Mech.\ Cat. \bf 135\/} (2022), 619--637.

\bibitem{bei23:112498}
{\rm A.~Beigzadeh, M.~Alabbad, D.~Liu, K.~Aljohani, K.~Hakimov, T.~A. Kashif,
  K.~Zanganeh, E.~Croiset,  \biband~ A.~Farooq}.
  \newblock {Reaction kinetics for high pressure hydrogen oxy-combustion in the presence of high levels of H$_2$O and CO$_2$}.
\newblock {\em Combust.\ Flame \bf 247\/} (2023), 112498.

\bibitem{dra87:471}
{\rm P.~Dransfeld \biband~ H.~{Gg. Wagner}}.
\newblock {Comparative study of the reactions of ${}^{16}$OH and ${}^{18}$OH with H${}^{16}$O$_2$}.
\newblock {\em Z.\ Naturforsch.\ A \bf 42\/} (1987), 471--476.

\bibitem{key88:1193}
{\rm L.~F. Keyser}.
\newblock {Kinetics of the reaction hydroxyl + hydroperoxo $\to$ water + oxygen from 254 to 382 K}.
\newblock {\em J.~Phys.\ Chem. \bf 92\/} (1988), 1193--1200.

\bibitem{atk89:881}
{\rm R.~Atkinson, D.~L. Baulch, R.~A. Cox, R.~F. {Hampson Jr.}, J.~A. Kerr,
  J.~Troe,  \biband~ R.~T. Watson}.
\newblock {Evaluated kinetic and photochemical data for atmospheric chemistry: Supplement III. IUPAC subcommittee on gas kinetic data evaluation for atmospheric chemistry}.
\newblock {\em J.~Phys.\ Chem. ~Ref. ~Data \bf 18\/} (1989), 881.

\bibitem{bau05:757}
{\rm D.~L. Baulch, C.~T. Bowman, C.~J. Cobos, R.~A. Cox, T.~Just, J.~A. Kerr,
  M.~J. Pilling, D.~Stocker, J.~Troe, W.~Tsang, R.~W. Walker,  \biband~
  J.~Warnatz}.
\newblock {Evaluated kinetic data for combustion modeling: Supplement II}.
\newblock {\em J.~Phys.\ Chem. ~Ref. ~Data \bf 34\/} (2005), 757--1397.

\bibitem{jpl11}
{\rm J.~B. Burkholder, S.~P. Sander, J.~Abbatt, J.~Barker, E.~L. Fleming,
  R.~Friedl, R.~Huie, C.~Jackman, C.~E. Kolb, M.~Kurylo, V.~Orkin,  \biband~
  P.~Wine}.
\newblock {\em NASA Data Evaluation: Chemical Kinetics and Photochemical Data
  for Use in Atmospheric Studies}.
\newblock JPL publication, {J}et {P}ropulsion {L}aboratory, {P}asadena, 2014.

\bibitem{kap02:4392}
{\rm C.~Kappel, K.~Luther,  \biband~ J.~Troe}.
\newblock {Shock wave study of the unimolecular dissociation of H$_2$O$_2$ in its falloff range and of its secondary reactions}.
\newblock {\em Phys.\ Chem.\ Chem.\ Phys. \bf 4\/} (2002), 4392--4398.

\bibitem{sri06:6602}
{\rm N.~K. Srinivasan, M.-C. Su, J.~W. Sutherland, J.~V. Michael,  \biband~
  B.~Ruscic}.
\newblock {Reflected shock tube studies of high-temperature rate constants for OH + NO$_2$ $\to$ HO$_2$ + NO and OH +HO$_2$ $\to$ H$_2$O + O$_2$}.
\newblock {\em J.~Phys.\ Chem.~A \bf 110\/} (2006), 6602--6607.

\bibitem{hon10:5520}
{\rm Z.~Hong, S.~S. Vasu, D.~F. Davidson,  \biband~ R.~K. Hanson}.
\newblock {Experimental study of the rate of OH + HO$_2$ $\to$ H$_2$O + O$_2$ at high temperatures using the reverse reaction}.
\newblock {\em J.~Phys.\ Chem.~A \bf 114\/} (2010), 5520--5525.

\bibitem{zha13:7381}
{\rm T.~Zhang, W.~Wang, C.~Li, Y.~Du,  \biband~ J.~L\"{u}}.
\newblock {Catalytic effect of a single water molecule on the atmospheric reaction of HO$_2$ + OH: fact or fiction? A mechanistic and kinetic study}.
\newblock {\em RSC Adv. \bf 3\/} (2013), 7381--7391.

\bibitem{bur13:1540}
{\rm M.~P. Burke, S.~J. Klippenstein,  \biband~ L.~B. Harding}.
\newblock {A quantitative explanation for the apparent anomalous temperature dependence of OH + HO$_2$ = H$_2$O + O$_2$ through multi-scale modeling}.
\newblock {\em Proc. Combust. Inst. \bf 34\/} (2013), 547--555.

\bibitem{hon13:565}
{\rm Z.~Hong, K.-Y. Lam, R.~Sur, S.~Wang, D.~F. Davidson,  \biband~ R.~K.
  Hanson}.
\newblock {On the rate constants of OH + HO$_2$ and HO$_2$ + HO$_2$: A comprehensive study of H$_2$O$_2$ thermal decomposition using multi-species laser absorption}.
\newblock {\em Proc. Combust. Inst. \bf 34\/} (2013), 565--571.

\bibitem{bur13:547}
{\rm M.~P. Burke, S.~J. Klippenstein,  \biband~ L.~B. Harding}.
\newblock {A quantitative explanation for the apparent anomalous temperature dependence of OH + HO$_2$ $=$ H$_2$O + O$_2$ through multi-scale modeling}.
\newblock {\em Proc.\ Combust.\ Inst. \bf 34\/} (2013), 547--555.

\bibitem{pal18:4478}
{\rm M.~Monge-Palacios \biband~ S.~M. Sarathy}.
\newblock {{\it Ab inition} and transition state theory of the OH + HO$_2$ $\to$ H$_2$O + O$_2$ (${}^3\Sigma_g^-$)/O$_2$(${}^1\Delta_g$) reactions: yield and role of O$_2$(${}^1\Delta_g$) in H$_2$O$_2$ decomposition and in combustion of H$_2$}.
\newblock {\em Phys.\ Chem.\ Chem.\ Phys. \bf 20\/} (2018), 4478--4489.

\bibitem{zha18:8152}
{\rm T.~Zhang, X.~Lan, Z.~Qiao, R.~Wang, X.~Yu, Q.~Yu, Z.~Wang, L.~Jin,
  \biband~ Z.~Wang}.
\newblock {Role of the (H$_2$O)$_n$ ($n=1-3$) cluster in the HO$_2$ + HO $\to$ ${}^3$O$_2$ + H$_2$O reaction: mechanistic and kinetic studies}.
\newblock {\em Phys.\ Chem.\ Chem.\ Phys. \bf 20\/} (2018), 8152--8165.

\bibitem{liu19:12667}
{\rm Y.~Liu, M.~Bai, H.~Song, D.~Xie,  \biband~ J.~Li}.
\newblock {Anomalous kinetics of the reaction between OH and HO$_2$ on an accurate triplet state potential energy surface}.
\newblock {\em Phys.\ Chem.\ Chem.\ Phys. \bf 21\/} (2019), 12667.

\bibitem{son20:134309}
{\rm Q.~Song, Q.~Zhang,  \biband~ Q.~Meng}.
\newblock {Revisiting the Gaussian process regression for fitting high-dimensional potential energy surface and its application to the OH + HO$_2$ $\to$ O$_2$ + H$_2$O reaction}.
\newblock {\em J.~Chem.\ Phys. \bf 152\/} (2020), 134309.

\bibitem{men18:8320}
{\rm Q.~Meng}.
\newblock {Ring-polymer molecular dynamics with coarse-grained treatment of the rate coefficients of chlorine atom reactions with methane, ethane, and propane}.
\newblock {\em J.~Phys.\ Chem.~A \bf 122\/} (2018), 8320.

\bibitem{liu20:3331}
{\rm Y.~Liu, H.~Song, D.~Xie, J.~Li,  \biband~ H.~Guo}.
\newblock {Mode specificity in the OH + HO$_2$ $\to$ H$_2$O + O$_2$ reaction: Enhancement of reactivity by exciting a spectator mode}.
\newblock {\em J.~Am.\ Chem.\ Soc. \bf 142\/} (2020), 3331--3335.

\bibitem{son24:597}
{\rm Q.~Song, X.~Zhang, Z.~Miao,  \biband~ Q.~Meng}.
\newblock {Construction of mode-combination Hamiltonian under the grid-based representation for the quantum dynamics of OH + HO$_2$ $\to$ O$_2$ + H$_2$O}.
\newblock {\em J.~Chem.\ Theory Comput. \bf 20\/} (2024), 597--613.

\bibitem{gat09:1}
{\rm F.~Gatti \biband~ C.~Iung}.
\newblock {Exact and constrained kinetic energy operators for polyatomic molecules: The polyspherical approach}.
\newblock {\em Phys.\ Rep. \bf 484\/} (2009), 1--69.

\bibitem{wan03:1289}
{\rm H.~Wang \biband~ M.~Thoss}.
\newblock {Multilayer formulation of the multiconfiguration time-dependent Hartree theory}.
\newblock {\em J.~Chem.\ Phys. \bf 119\/} (2003), 1289--1299.

\bibitem{man08:164116}
{\rm U.~Manthe}.
\newblock {A multilayer multiconfigurational time-dependent Hartree approach for quantum dynamics on general potential energy surfaces}.
\newblock {\em J.~Chem.\ Phys. \bf 128\/} (2008), 164116.

\bibitem{ven11:044135}
{\rm O.~Vendrell \biband~ H.-D. Meyer}.
\newblock {Multilayer multiconfiguration time-dependent Hartree method: Implementation and applications to a Henon-Heiles Hamiltonian and to pyrazine}.
\newblock {\em J.~Chem.\ Phys. \bf 134\/} (2011), 044135.

\bibitem{wan15:7951}
{\rm H.~Wang}.
\newblock {Multilayer multiconfiguration time-dependent Hartree theory}.
\newblock {\em J.~Phys.\ Chem.~A \bf 119\/} (2015), 7951.

\bibitem{wer12:242}
{\rm H.-J. Werner, P.~J. Knowles, G.~Knizia, F.~R. Manby,  \biband~
  M.~Sch{\"u}tz}.
\newblock {Molpro: A general-purpose quantum chemistry program package}.
\newblock {\em WIREs Comput. Mol. Sci. \bf 2\/} (2012), 242--253.

\bibitem{molpro}
{\rm H.-J. Werner \biband~ P.~J. Knowles}.
\newblock MOLPRO is a package of ab initio programs. Further information can be obtained from https://www.molpro.net.

\bibitem{wer85:5053}
{\rm H.-J. Werner \biband~ P.~J. Knowles}.
\newblock {A second order MCSCF Method with optimum convergence}.
\newblock {\em J.~Chem.\ Phys. \bf 82\/} (1985), 5053--5063.

\bibitem{kno85:259}
{\rm P.~J. Knowles \biband~ H.-J. Werner}.
\newblock {An efficient second order MCSCF method for long configuration expansions}.
\newblock {\em Chem.\ Phys.\ Lett. \bf 115\/} (1985), 259--267.

\bibitem{kno84:315}
{\rm P.~J. Knowles \biband~ N.~C. Handy}.
\newblock {A new determinant-based full configuration interaction method}.
\newblock {\em Chem.\ Phys.\ Lett. \bf 111\/} (1984), 315--321.

\bibitem{dun89:1007}
{\rm T.~H. {Dunning, Jr.}}
\newblock {Gaussian basis sets for use in correlated molecular calculations I. The atoms boron through neon and hydrogen}.
\newblock {\em J.~Chem.\ Phys. \bf 90\/} (1989), 1007.

\bibitem{ken92:6796}
{\rm R.~A. Kendall, T.~H. Dunning,  \biband~ R.~J. Harrison}.
\newblock {Electron affinities of the first-row atoms revisited. Systematic basis sets and wave functions}.
\newblock {\em J.~Chem.\ Phys. \bf 96\/} (1992), 6796--6806.

\end{thebibliography}

\end{document}